**Rate-Programmable Ionic–Redox Switching with Tunable Volatility in $CuCrP_2S_6$**

*Suzanne Lancaster[1*], Francesco Calavalle[2], Mayank Sharma[1,3], Lucía Olano-Vegas[1,3], Garen Avedissian[1,4], Tanweer Ahmed[1], Marco Gobbi[5,6], Beatriz Martín-García[1,5,6], Beatrice Fraboni[2,7], Fèlix Casanova[1,6], Luis E. Hueso[1,6]*

1 – CIC NanoGUNE BRTA, 20018 Donostia-San Sebastián, Basque Country, Spain

2 - Department of Physics and Astronomy, University of Bologna, Viale Berti Pichat 6/2, Bologna, 40127 Italy

3 - Departamento de Polímeros y Materiales Avanzados: Física, Química y Tecnología, University of the Basque Country (UPV/EHU), Donostia-San Sebastian, 20018, Spain

4 - Donostia International Physics Center, DIPC, 20018 Donostia-San Sebastian, Basque Country, Spain

5 - Materials Physics Center CSIC-UPV/EHU, 20018 Donostia-San Sebastián, Spain

6 - IKERBASQUE, Basque Foundation for Science, 48009 Bilbao, Spain

7 - INFN, Sezione di Bologna, Viale Berti Pichat 6/2, Bologna, 40127 Italy

**Abstract:** Metal thiophosphates are emerging as a multifunctional material platform for neuromorphic electronics due to their accessible polar phases and ion dynamics on biologically relevant timescales. While resistive switching in these materials is frequently attributed to ferroelectric or antiferroelectric polarization, the intrinsic role of ion dynamics remains underexplored. Here, we isolate and demonstrate purely ion-driven resistive switching in paraelectric $CuCrP_2S_6$. Robust and reproducible resistive switching is observed in the absence of measurable ferroelectricity. The conductance can be tuned through both voltage amplitude and sweep rate, revealing a rate dependence characteristic of ion dynamics. The resulting resistance states exhibit controllable volatility, where switching rate determines the decay time constant of the readout current, attributed to ionic relaxation. Using either inert or reactive electrodes, we observe electrical evidence of solid-state redox activity associated with the interfacial reduction of native $Cu^+$ ions, enabling controlled formation of filamentary conduction pathways. Analysis of this process allows extraction of the $Cu^+$ diffusion coefficient, providing quantitative insight into the underlying transport kinetics. The understanding of ionic-redox based resistive switching in $CuCrP_2S_6$ is crucial for unleashing its full potential as a material platform for dual- or multi-mode operation.

**Introduction**

Neuromorphic computing is an emerging paradigm which mimics biological functions for information processing [1], where artificial devices are used to recreate synaptic and neuronal functions of the human brain[2]. Neuromorphic processing leverages the energy efficiency, parallelism and nonlinear dynamics of novel devices in order to perform tasks for which the human brain is particularly well-suited, including image recognition, sensing, motor control, and learning[3,4]. Such devices operate over a wide range of biologically relevant timescales. In particular, switching due to slow ion dynamics allows for real-time information processing in response to environment stimuli[4,5], while devices operating across different timescales, combining multiple resistive switching mechanisms and tunable retention times, allow a versatile emulation of real biological systems[6,7]. Since the requirements for different aspects of a neuromorphic processor, for example, learning or memory, can often conflict, there is strong interest in materials which show different properties and independent switching mechanisms when operated at different timescales or voltage ranges[8,9].

Particularly promising are the metal thiophosphates $CuInP_2S_6$ (CIPS) and $CuCrP_2S_6$ (CCPS), where a wealth of field-driven electrical and ionic effects are known to co-exist[10,11]. Increasing the magnitude of the applied field or the temperature can drive the material from a ferroelectric switching regime, into ion 'hopping' between van der Waals (vdW) layers and eventually filament formation[11]. In both CIPS and CCPS the presence of ionic migration has been evidenced electrically[12,13] as well as by the formation and dissolution of topographical surface features in scanning probe microscopy experiments [14,15]. In CIPS, both polar ordering and ion migration have been observed in aberration-corrected Scanning Transmission Electron Microscopy (STEM)[16] and energy dispersive X-ray spectroscopy[17].

For many decades, polar ordering has been known to exist in metal thiophosphates[18,19], but recent extensive research into two-dimensional (2D) CIPS and CCPS has been partly driven by the rising success and industrial interest in other ferroelectric materials such as $Hf_{0.5}Zr_{0.5}O_2$ (HZO) and AlScN[20,21] as well as emergent ferroelectricity in other 2D materials systems[22,23]. While many reports exist showing room temperature ferroelectricity in CIPS[24–26], CCPS has been less investigated and results are inconsistent, with ferroelectric, antiferroelectric and paraelectric phases all reported at room temperature[27–29]. Similar to in other materials, routes exist for controlling the polar ordering in the thiophosphates[30], and the strong interplay of ionic migration and ferroelectricity[31] open up a wealth of possibilities for driven ferroelectricity in these and related materials[32]. However, isolating the role of ionic migration in resistive switching is difficult, and has so far only been performed in CIPS[10]. While redox-based resistive switching often requires an active electrode, switching with native cations is forming-free: work on doped diffusive memristors[33] and more recently, intrinsic-ion memristors[34] have been explored, and the latter has shown that exploiting native cations for resistive switching can simplify processing while improving crystal quality.

Vertical devices based on CCPS have been operated at voltages < 1.2 V, making it ideal for low-power devices integrated in the BEOL[35] (here, resistive switching was attributed to antiferroelectric-to-ferroelectric transitions). However, since CCPS bulk growth requires high temperatures[36], large-scale CMOS integration would more likely rely on scalable transfer techniques[37]. Nonetheless, a better understanding of the ionic behaviors in CCPS and electrochemical behavior at the electrodes is an essential step to understand their suitability and reliability for integration in devices, and to disentangle different mechanism contributions to their device behavior.

In this paper we investigate the electrical response of two-terminal devices based on paraelectric CCPS and demonstrate that volatile resistive switching behavior can be engineered by DC voltage sweeps. Combined with Piezoresponse Force Microscopy (PFM) and Kelvin-Probe Force Microscopy (KPFM), the resulting resistive switching at a frequency of ≤ 2 Hz is attributed solely to interlayer hopping and interfacial electrochemistry. The switching pulse width can be used to control not only the conductivity, but also the volatility of the device state. By driving the devices into near-breakdown, redox reactions and filament formation are electrically observed, eventually causing irreversible structural changes as evidenced by Raman spectroscopy. This work highlights the critical role on macroscopic device behavior of $Cu^+$ reduction and oxidation at the electrode interfaces, provides insight into the ion dynamics and electrochemistry which allow access to biologically relevant timescales with CCPS, and highlights the need for carefully controlled operation of CCPS in resistive switches.

## Results and Discussion

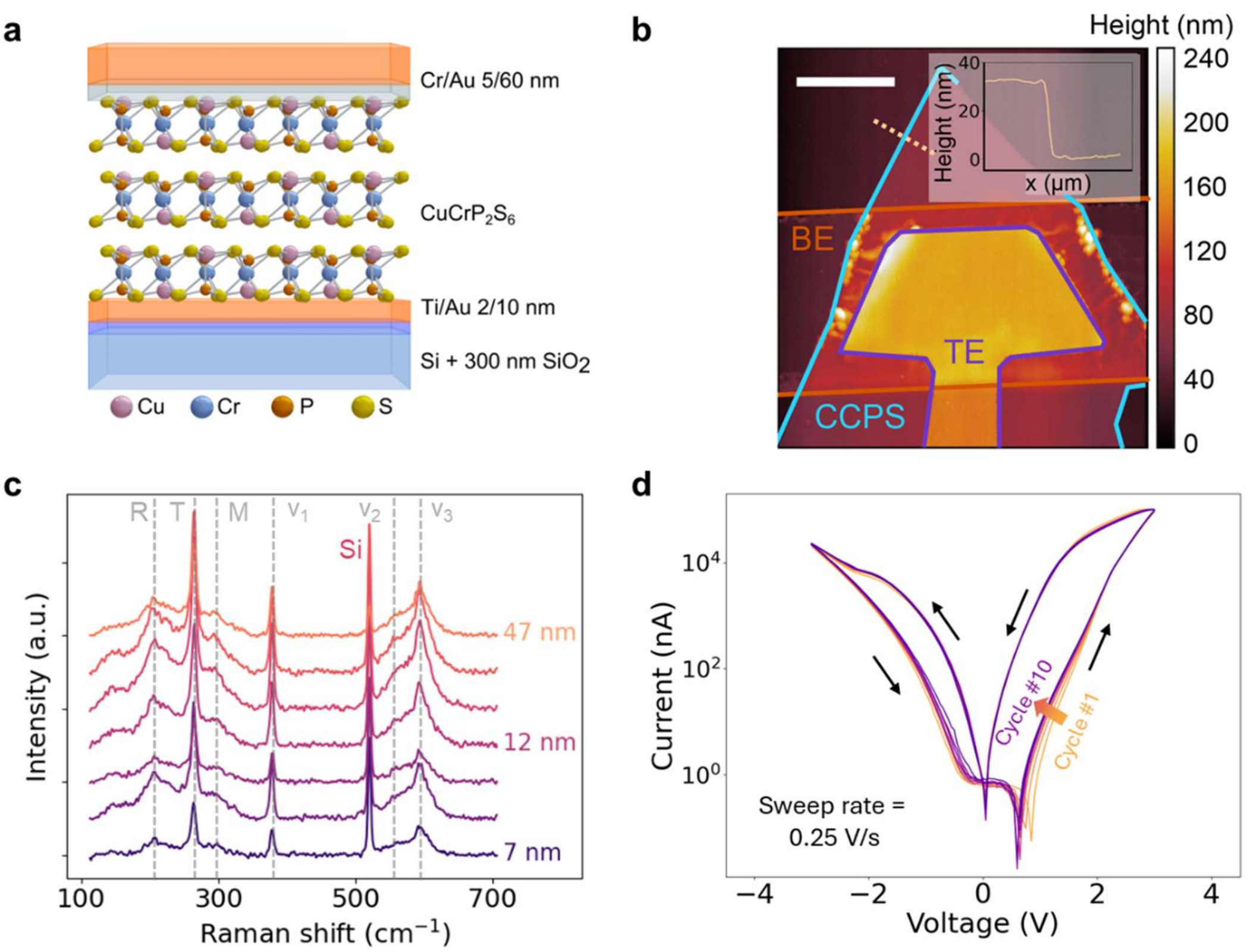

Figure 1: Basic characterization of CCPS-based resistive switching devices. (a) schematic of CCPS-based vertical device. (b) Atomic force micrograph of Device 1 (scale bar is 2 μm). The inset shows a typical linescan used to determine flake thickness. (c) Raman spectra measured on flakes of CCPS exfoliated on Au, as a function of flake thickness. (d) Typical resistive switching curves of Device 1, obtained over ten positive/negative voltage cycles at 0.25 V/s. The black arrows indicate the sweep direction.

We first discuss the basic characterization of the vertical devices used in this work. The device structure is shown in Figure 1a,b. The devices consist of CCPS layers sandwiched between a thin Ti (2 nm)/Au (10 nm) bottom electrode (BE) and the top electrode (TE) is chosen as Cr (5 nm)/Au (50 nm). In resistive switching devices, it can be expected that the use of one inert ("blocking") and one reactive ("active") electrode will increase the *On*/*Off* ratio. The atomic force microscopy (AFM) map in Figure 1b can be used to verify the device structure and the thickness of the measured flake, which in the case of this Device 1, shown here, is 33 nm (see extracted linescan inset in Figure 1b). A summary of all other devices analyzed in this paper is presented in Table S1, and AFM images for other CCPS flakes shown in the manuscript are presented in Figure S1.

Plotted in Figure 1c are Raman spectra measured on flakes exfoliated from bulk single-crystal CCPS onto a thin blanket Au electrode. Several flakes of different thicknesses were mapped in order to identify whether phase changes with thickness or other identifiable peak shifts can be seen, which may indicate strain or structural disorder[38]. The peak positions agree well with bulk references under ambient pressure at room temperature[38] and show no change with thickness. The resolvable modes are the rotational (R) and translational (T, M) motion of the S-P-S bonds and the stretching vibrations: P-P ($\nu_1$) and P-S ($\nu_2$, $\nu_3$).

Figure 1d shows a set of typical resistive switching curves measured on Device 1. When sweeping in positive polarity, the device starts in a low-conductance state, and the device current increases when a positive voltage is applied on the TE. When switching with a negative voltage on the TE, the device conductance remains high initially, but reduces on the return sweep, returning to the low conductance state. From here on, positive voltage pulses will be referred to as the *Set* pulse, while negative voltage pulses will be referred to as *Reset*. It should be noted that in contrast to many previous reports, this resistive switching occurs in the absence of ferroelectricity, which was ruled out in our material by PFM measurements (Figure S2). The resistive switching is stable over multiple switching cycles, as shown in Figure 1d.

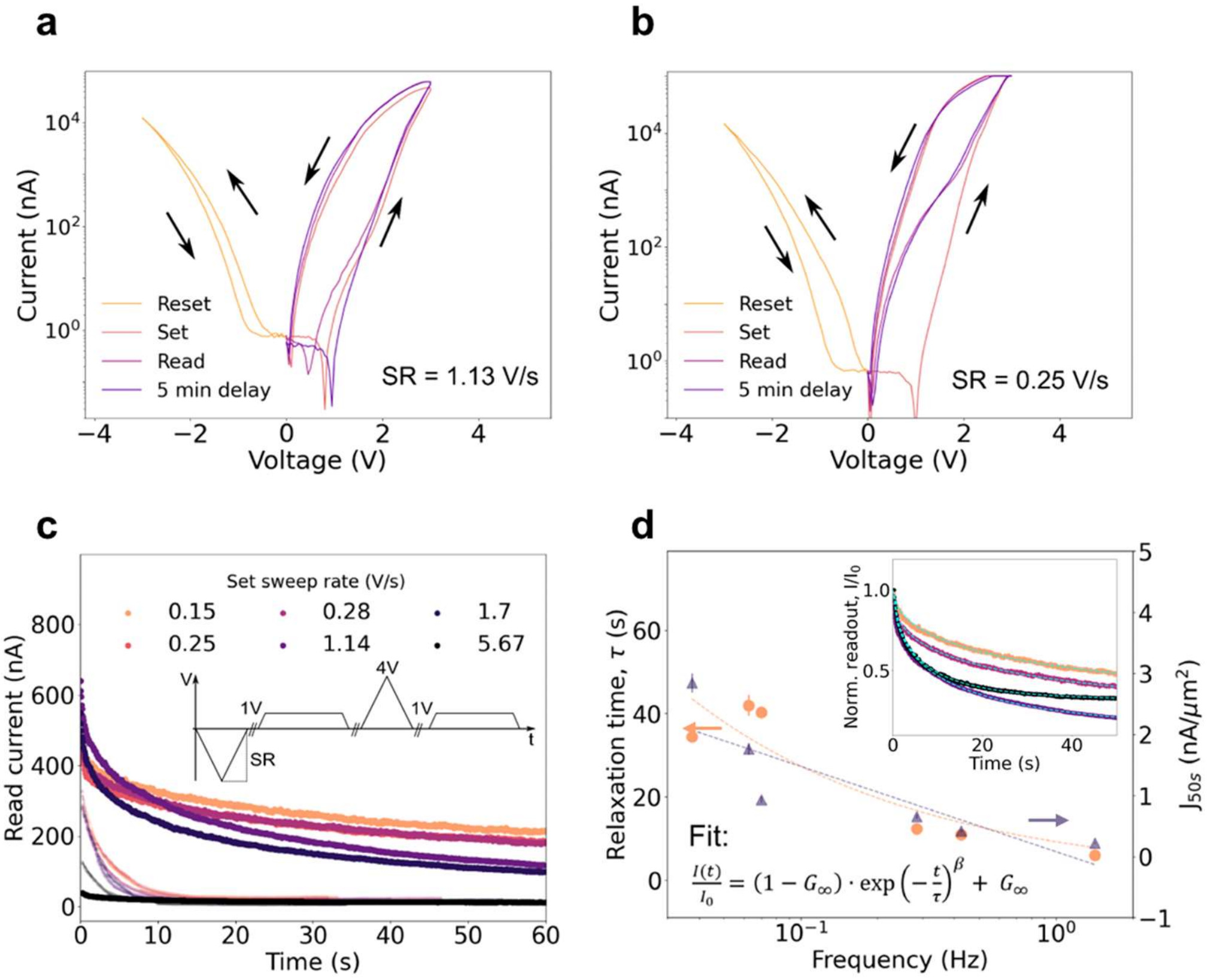

Figure 2: Rate-based resistive switching behavior of Cr/CCPS/Au vertical device. (a) Resistive switching at a sweep rate of 1.13 V/s, with a delay of 0 s and 5 mins after programming (black arrows represent the direction of resistive switching). (b) Resistive switching at a reduced sweep rate of 0.25 V/s, with a delay of 0 s and 5 mins after programming. (c) Readout current of device state, using a 1 V read pulse, after programming with pulses of 4 V amplitude and varying sweep rates. Bold traces represent readout after *Set* pulses, while light traces represent readout after *Reset*, and the applied electrical measurement scheme is shown inset (performed at 0.12 mbar in air). (d) Relaxation times τ (orange circles) extracted from the readout data and current density at 50 s (purple triangles), as a function of programming frequency. The inset shows the normalized readout current which was fitted with Eqn. 1 (also shown inset).

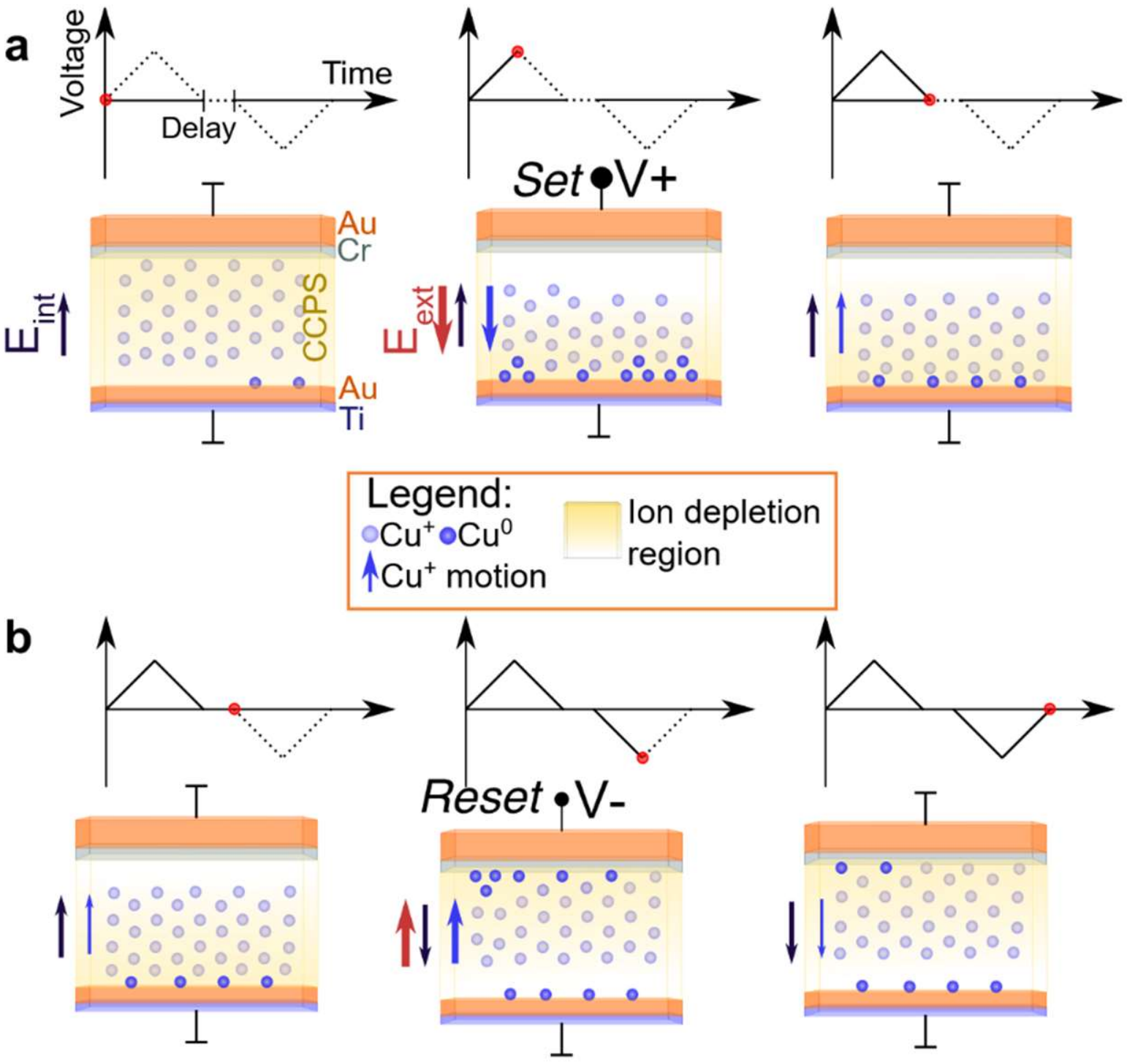

Figure 3: Model of resistive switching in CCPS-based ionic-redox switching devices upon application of (a) positive (*Set*) and (b) negative (*Reset*) pulses.

Resistive switching devices intended for neuromorphic computing require multiple resistance states which can be accessed by changing the pulse width, amplitude or in the best case, by repeated identical pulses[39]. Synaptic plasticity and short- and long-term potentiation can be mimicked through an interplay of programming parameters and temporal dynamics[2,6]. *Figure 2*a,b shows how the pulse sweep rate (pulse width) affects resistive switching in our CCPS-based devices. The devices were first switched to the *Set* state at a chosen sweep rate, followed immediately by a second triangular pulse to measure any change in conductivity. A further *Read* measurement was then taken after 5 minutes to determine whether the change was non-volatile.

At the faster sweep rate of 1.13 V/s (equivalent to a triangular pulse duration of ~5 s; measurement details are provided in Figure S3), only a small increase in conductance was observed, mainly below 2 V. In the initial device state, the current does not change sign until approximately +1 V, indicating the presence of an internal bias field caused by $Cu^+$ being preferentially shifted towards the top electrode (TE). On the return sweep, the current changes sign at 0 V, indicating that the internal bias field has been screened by mobile charges. On the other hand, during a 5 minute delay time, the device returns almost completely to its initial state, indicating the change in the conductance was highly volatile.

In contrast, reducing the sweep rate to 0.25 V/s (~24 s triangular pulse; Figure 2b) produces clear remanent effects in the resistive switching behavior. After the initial *Set*, the current exhibits a strong increase in conductance, and after 5 minutes only partially relaxes towards the initial state. During the subsequent *Reset*, the conductance remains high and the initial state is not fully recovered. This is further confirmed by a final *Set* pulse (Figure S4b), where the current remains higher than on the initial sweep. These results show that slower pulses induce more persistent changes in the device state. Additional sweep rates (Figure S5) further demonstrate access to a broad range of conductance states depending on the programming conditions.

In *Figure 2*c the readout of the resistance state is performed not with additional slow triangular pulses but with a single square 'Read' pulse, as shown inset and in Figure S3. A readout voltage of 1 V is chosen to maximize the *On*/*Off* ratio, which can be tuned between 2 for a sweep rate of 5.67 V/s to 16 for 0.15 V/s. The bold curves represent the readout current in the *On*-state and they show a conductance which depends on the *Set* sweep rate. The light curves represent the readout current in the *Off*-state and they quickly tend to a low current value, as does the current after a *Set* pulse at 5.67 V/s. Inset in *Figure 2*d the normalized readout current is plotted against readout time for the *On*-state. It can clearly be seen that the sweep rate controls the volatility, evidenced by the varying slope of the normalized readout curve. These were fitted with a stretched exponential function, describing ionic relaxation[40–42]:

$$\frac{I(t)}{I_0} = (1 - G_\infty) \cdot \exp\left(-\frac{t}{\tau}\right)^{\beta} + G_\infty \quad (1)$$

Where $I(t)/I_0$ is the normalized readout current, $G_\infty$ is the expected steady-state normalized conductance at t → ∞, t is the readout time, τ is a time constant which here describes the ionic relaxation dynamics and β is an exponential factor which describes the disorder in the system. A lower β corresponds to a higher disorder, and in this case, β was found to be around 0.5-0.6 at all sweep rates. This suggests a broad distribution of relaxation times and strongly heterogeneous ionic relaxation with multiple pathways, consistent with $CuCrP_2S_6$ acting as a disordered solid electrolyte[43].

The extracted values of τ are plotted in *Figure 2*d (orange circles). The value of τ decreases with an increasing frequency, indicating that there is a significant, fast ionic relaxation component to the readout current which is strongly frequency-dependent. On the other hand, the purple triangles represent a longer acting change in the conductance state, represented by $J_{50s}$, which is the current density after 50 s of measurement, for each state. The data also show a strong frequency dependence, whereby the conductance of the device decreases logarithmically with sweep frequency. τ and $J_{50s}$ were extracted for multiple measurements on three different devices and all data show the same qualitative trend, as shown in Figure S6.

The schematics in Figure *3*: Model of resistive switching in CCPS-based ionic-redox switching devicesFigure 2 illustrate the proposed mechanism behind the resistance change behavior, which is attributed to the combined effects of ion migration and interfacial electrochemistry. Previous studies[13,15,16], have shown that the native $Cu^+$ ions in CIPS and CCPS can migrate across the vdW gaps under high electric fields (0.05 V/nm for CIPS[13]) and/or prolonged electrical pulses[14].

Under positive bias at the Cr electrode, $Cu^+$ ions migrate towards the bottom electrode (BE), reducing the ion depletion at the BE and increasing conductance. The accumulation or depletion of $Cu^+$ strongly impacts the conduction and modulates the interfacial energy barriers[44], and the oxidation state of Cr will also impact the contact made to CCPS. Considering that under equilibrium conditions, hole conduction dominates in CCPS[36], the initial presence of $Cu^+$ near the TE may cause band bending which limits hole injection. However, it is not possible from our measurements to know which barrier is limiting, and indeed, the dynamic ionic system means that the limiting barrier may change during switching. Under fast sweep rates, the conductivity change is strongly volatile. During the delay time, $Cu^+$ gradually drifts back towards the TE[45] as ions relax back to their equilibrium positions. Applying a *Reset* pulse (negative bias at the TE) drives $Cu^+$ back towards the TE, restoring the initial, low conductance state. When the *Set* pulse duration is increased, the longer action of the electric field allows more extensive $Cu^+$ migration across the vdW gaps. This increases both the number of migrating ions and the stability of their redistributed positions, thereby widening the resistive switching memory window and leading to a longer-lasting change in conductance.

The schematics additionally indicate reduction of $Cu^+$ at the interfaces, which is discussed in more detail later. Partial removal of $Cu^+$ from the lattice may stabilize ion migration, since multiple $Cu^+$ occupations within the same unit cell are metastable and have only been observed under electron beam irradiation[16]. Overall, resistive switching in CCPS arises from the combination of short-term ionic processes, described by the relaxation time constant $\tau$, and longer lasting electrochemical changes, reflected by the conductance term $J_{50s}$. The interaction between these mechanisms produces tunable volatility controlled by the pulse conditions, providing a route towards synaptic functionality in CCPS.

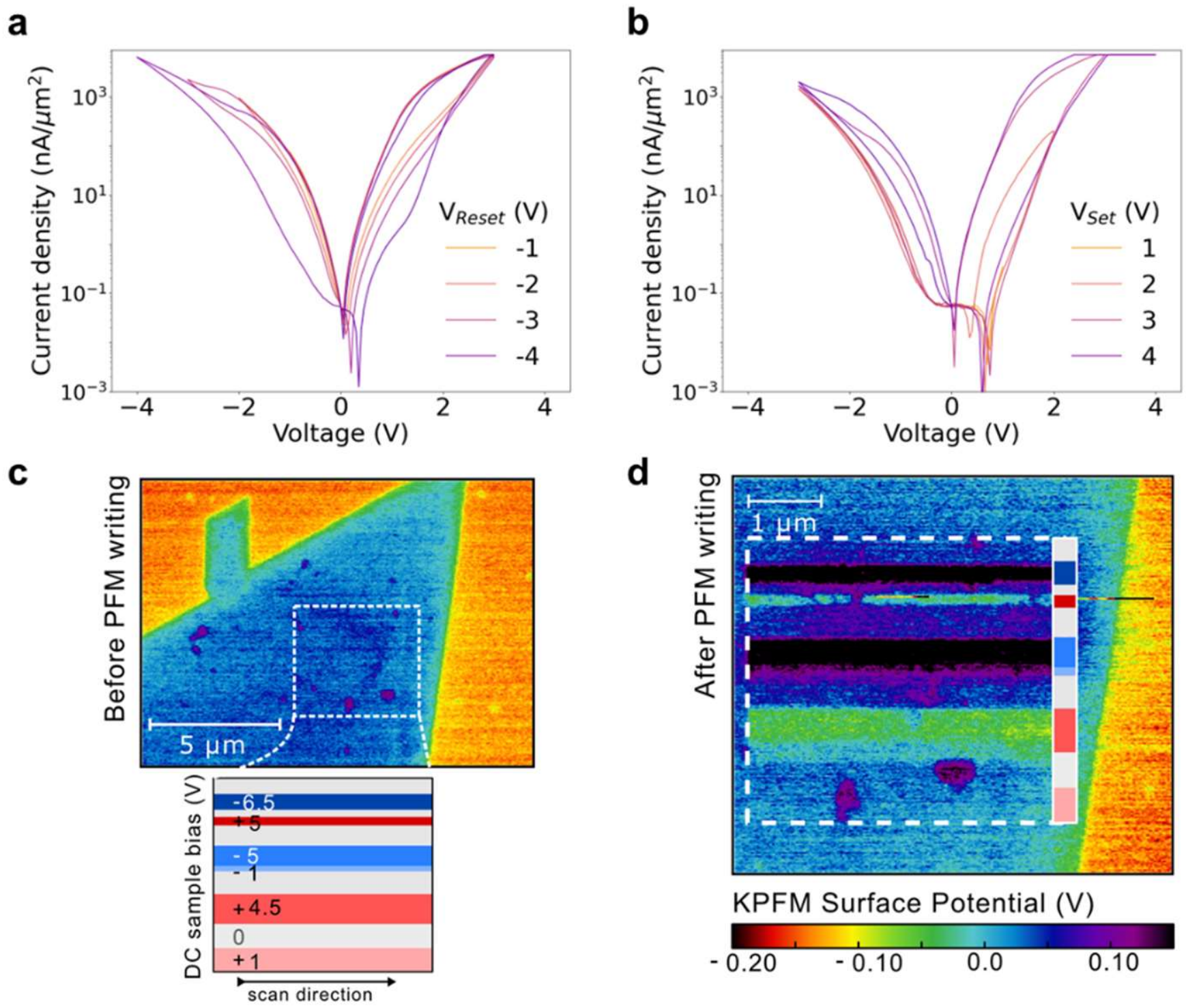

Figure 4: Voltage dependence on resistive switching. (a) Resistive switching curves when fixing the *Set* voltage at +3 V and varying the *Reset* voltage between -0.5 and 4 V, and (b) when fixing the *Reset* Voltage at -3 V and varying the *Set* voltage between 0.5 and 4 V. Sweep rate for (a) and (b) is constant at 0.25 V/s. (c) Flake surface before PFM (above) and writing scheme at 0.5 Hz (below) on a CCPS flake. (d) KPFM map of the written stripes. The inset shows a visual guide to the writing voltages indicated in (c). Electrical measurements were performed in air at 0.12 mbar and KPFM was performed in Ar.

Further experiments showed that the resistive switching behavior and remanent surface potential changes are consistent with voltage-dependent $Cu^+$ redistribution. Figure 44 shows the impact of applied voltage on resistive switching, when fixing the sweep rate, (for clarity, only some voltages are plotted here; full curves at steps of 0.5 V are shown in Figure S7).

In Figure 44a, first a *Set* pulse of +3 V is applied to the device. Variable *Reset* pulses from −0.5 V to −4 V are then applied, with +3 V *Set* pulses inserted between them. This allows the evolution of the positive-polarity hysteresis to be monitored as the *Reset* voltage increases. For $V_{reset} > -1.5$ V only a narrow hysteresis is observed, indicating that the initial pulse increases the conductivity and the device remains in the high-conductance state. When $V_{reset} \leq -1.5$ V, the conductance begins to decrease during the subsequent positive sweep and the hysteresis loop opens, showing that the *Reset* pulse now affects the device state. The hysteresis width eventually saturates for V < 3.5 V.

The pulse polarity is reversed in Figure 44b, and the opposite behavior is observed. At low $V_{set}$, the conductance during the negative sweep remains low. However, for $V_{set} > 2$ V, the conductance increases during the following negative sweep and the hysteresis loop again opens. In both polarities, threshold voltages are observed ($V_{reset} \leq -1.5$ V and $V_{set} \geq 2$ V) and the hysteresis loops saturate above |3.5 V|. The threshold voltages correspond to a field across the CCPS of 0.045 – 0.06 V/nm, which is on the order of the field for ionic domination found in CIPS[13].

Figure 44c, d show Kelvin Probe Force Microscopy (KPFM) experiment on a CCPS flake of 17-30 nm thickness on Au, in Ar atmosphere. Measurements were performed on a 17-30 nm thick CCPS flake on Au under Ar atmosphere. Figure 4c shows the flake before switching (top panel) and the DC voltages ($V_{DC}$) applied to the bottom electrode during scanning at 0.5 lines/s (bottom panel). Afterwards, KPFM measurements were performed under $V_{AC}$ = 1 V to detect remanent changes in surface potential (Figure 44d). No measurable surface potential changes were observed for voltage scans of ±1 V. However, at higher voltages, the written patterns from Figure 4c became clearly visible in the KPFM maps. Under positive $V_{DC}$, the surface potential decreases relative to the background, indicating accumulation of positive charge near the top interface. Under negative $V_{DC}$, the surface potential increases, suggesting $Cu^+$ depletion. The difference in surface potential

when the sample is poled at -6.5 V and +5 V is 200 mV, comparable to values reported for CIPS during lateral $Cu^+$ transport[46] where $Cu^+$ migration was corroborated with Raman spectroscopy. These experiments were performed in Ar atmosphere; in ambient conditions, surface adsorbates rapidly screen the internal charges.

Overall, the KPFM results support the switching mechanism proposed in *Figure 3*: Model of resistive switching in CCPS-based ionic-redox switching devices upon application of (a) positive (Set) and (b) negative (Reset) pulses., indicating that the resistance changes originate from $Cu^+$ migration and the resulting modification of the local work function, which influences charge injection at the electrodes. The results additionally support the threshold voltages observed in Figures 4a,b, which likely correspond to an energy barrier for ion migration across the vdW gaps and therefore depend on both the applied voltage and pulse duration.

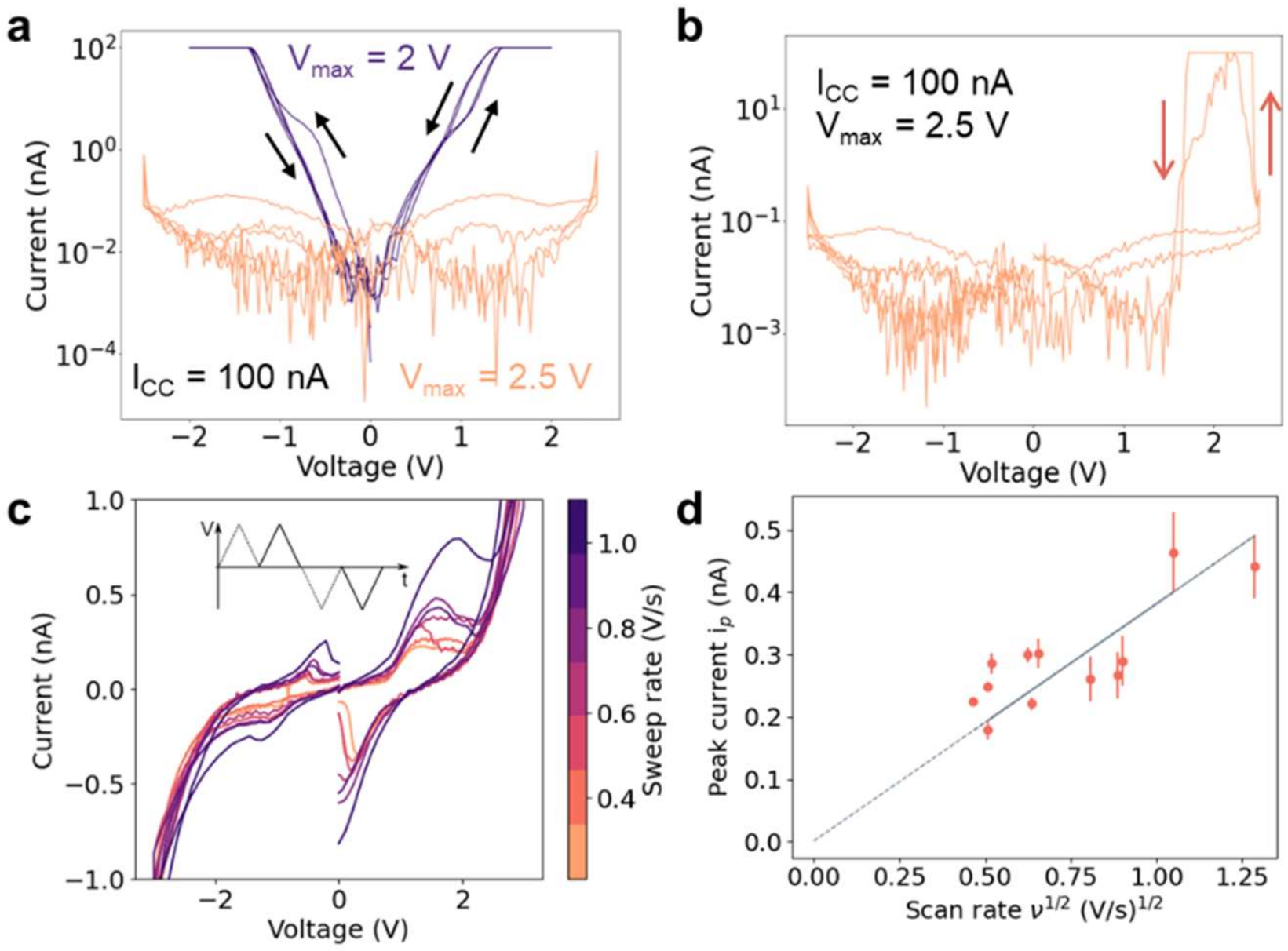

Figure 5: Evidence of solid-state redox activity in CCPS. (a) Transition from resistive switching regime to soft breakdown in Device 2. (b) Filament formation in the soft breakdown regime. (c) Plotting of Device 3 redox currents in the soft breakdown regime (measured with the pulse scheme

shown inset); (d) the peak heights in the positive polarity are extracted and plotted against scan rate $v^{1/2}$, showing a linear relationship. The data were measured in ambient atmosphere.

When measuring on some devices in ambient atmosphere, a soft-breakdown behavior was observed upon increasing the maximum voltage, as depicted in Figure 5a (Device 2). Within this regime, the device could be repeatably cycled (see Figure S8) and filamentary behavior could be switched on by applying repeated pulses in the same polarity, with an example shown in Figure 5b (this behavior is also seen in Figure S8). When applying the pulse train shown inset in Figure 5c (Device 3), a distinctive current response was observed on each pulse, which we isolated by taking only the second pulse in each voltage polarity. When different sweep rates are plotted together (Figure 55c), these currents show a clear signature of redox activity, which we attribute to the reduction and re-oxidation of $Cu^+$ at the Au and Cr electrodes for positive and negative applied voltages, respectively. To further verify this, the peak heights for the forward peaks at positive voltages are extracted and plotted in Figure 55d.

According to the Randles-Sevcik law[47,48], reversible or quasi-reversible redox reactions at room temperature follow the relationship,

$$i_p = 2.69 \cdot 10^5 n^{\frac{3}{2}} A C \sqrt{D v} \quad (2)$$

Where $i_p$ is the current peak height, $n$ is the number of electrons transferred in the reaction (here assumed to be 1), $A$ is the electrode area, $C$ is the molar concentration of reacting ions, $D$ is the diffusion coefficient of cations, and $v$ is the sweep rate. The linear fit in Figure 55d shows that the current peaks follow $i_p \propto \sqrt{v}$, with an intercept of 0 as expected. Therefore, the peaks are confirmed to arise from a redox reaction, and moreover, this relationship confirms that the peaks specifically arise from a freely diffusing species (rather than surface-bound), i.e., from $Cu^+$ which migrates through the lattice to the electrodes, and the process is (quasi-)reversible. For the Cr/CCPS/Au system, the redox peaks are bipolar, that is, a distinct reduction and re-oxidation peak can be observed in each direction, which we attribute to the reduction and re-oxidation of $Cu^+$ at the Au electrode and at the Cr electrode in the positive and negative voltage polarities, respectively, with the half-reaction $Cu^+ + e^- \rightleftharpoons Cu^0$. In classical cyclic voltammetry, the lowest potential reaction involving Cu is the half-reaction $Cu^{2+} + e^- \rightleftharpoons Cu^+$; we note that this reaction may be less favorable here due to the presence of native $Cu^+$ which likely has a strong impact on the reaction potentials, and the fact that $Cu^{2+}$ was found to be less energetically stable than the formation of Cu vacancies in CIPS[15]. From equation (2) the diffusion coefficient of $Cu^+$ in CCPS can be calculated (full details in supplementary Table S2), which yields a value of $3.65e^{-12}$ $cm^2/s$. Based on ionic conductivity data in literature[49] we can calculate a diffusion coefficient of $5.00 \times 10^{-12}$ $cm^2/s$, agreeing well with our measured value. Comparing this to CIPS, as calculated from the same paper ($D = 2.1 \times 10^{-13}$ $cm^2/s$) CCPS hosts significantly faster $Cu^+$ diffusion, making it a stronger candidate for resistive switches based on ion dynamics. It should be noted that the CIPS values calculated from reference[49] are also corroborated by calculations based on recent additional references[50,51] (full details can be found in Supplementary Table S3).

Since the reduction of $Cu^+$ at one interface requires oxidation at the opposite interface to maintain neutrality, this suggests either the presence of hydroxyl species at the electrodes due to processing residuals[52] or that lattice oxidation occurs at the opposite interface[15], and we observed a Raman signature consistent with lattice oxidation, as discussed later. We note that the redox peaks shown in Figure 55c are larger in the positive polarity when $Cu^+$ accumulates near Au. On the one hand, inert metals have been shown in electrochemical metallization cells (ECMs) to better support redox-based switching behavior[53], when compared to active metals which may form an interfacial oxide layer (e.g. Ta or in this case, Cr). On the other hand, the complementary lattice oxidation may be favorable at the Cr/CCPS interface due to a higher reaction of Cr with the sulfur framework, as shown via DFT simulations for other 3$d$ transition metals, Cu and Ni[54]. Therefore, a combination of these effects may explain the more effective redox reaction (and larger redox peaks) when a positive voltage is applied in this system.

Besides ion migration, it has been reported that CIPS can support a filamentary regime[11]. Here, the formation of filaments observed in CCPS can be directly attributed to the interfacial redox path discussed above. The filaments formed in our experiments were volatile, dissolving during the $Cu^0 - e^- \rightharpoonup Cu^+$ half-reaction on the ramp down (Figure 55b and Figure S8), and could occur in either polarity. Filaments could even be observed in both voltage polarities during a single measurement on the same device (Device 4, Figure S9), which indicates that the system is similar to bipolar ECMs reported in literature[55] with native $Cu^+$ cations; after filament dissolution near the electrode, much of the filament remains, increasing the probability of filament formation on the next pulse. However, once the filamentary regime is reached, filament formation can still be prevented by additional device cycling at lower voltages, which is depicted in Figure S10 and indicates that low-voltage cycling can essentially reset the device through the redistribution of $Cu^+$. Note that this filament dissolution could also be a time-dependent effect (i.e. the filament could be volatile, and dissolve over time) which was not investigated further in this study.

Measurements on a sample with Pt top electrode (Device 5) under low vacuum (0.12 mbar of air) also show redox-like peaks under similar soft-breakdown conditions, indicating that the redox reactions are inherent to the material system, occurring even with inert electrodes (Figure S11). This further confirms that native ions, i.e. $Cu^+$, must be the root of the redox behavior. Finally, in ferroelectric CIPS we observed the co-existence of filamentary and resistive switching behavior (Figure S12) hinting that redox occurs reversibly during the resistive switching regime, although the electrical signature could only be isolated when the device was in the soft-breakdown state. The appearance of filamentary switching in the resistive switching regime is a mechanism which could give rise to short- and long-term potentiation in these materials.

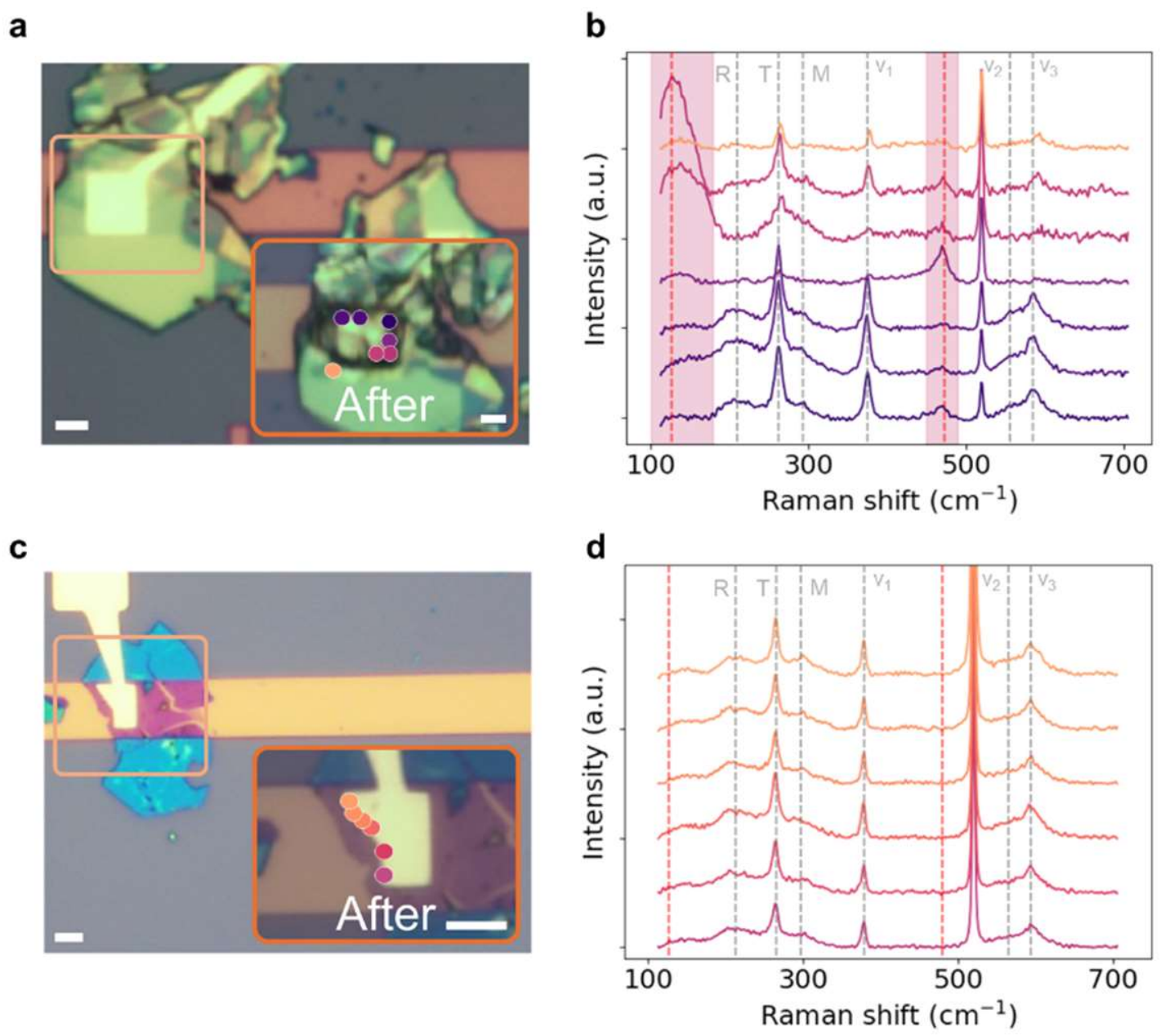

Figure 6: Signatures of irreversible redox in CCPS. (a) Optical microscope images of Device 3 before and after measurements, and (b) Raman spectra measured around the edge of the top electrode. (c) Optical microscope images of Device 6 before and after measurements, and (d) Raman spectra measured around the edge of the electrode. Scale bars are 3 µm. The colored dots inset in (a) and (c) indicate the approximate measurement position for the correspondingly colored spectra in (b) and (d).

Previous reports have shown that repeated polarization cycling of CIPS can lead to the extraction of $Cu^{+}$ from the lattice and a subsequent restructuring of S atoms, leading to a distinctive $S_8$ Raman signal[56]. As shown in Figure 6a,b, in the case where the devices showed the soft-breakdown behavior described above, clear topographical changes could be observed. Previous literature on conductive-AFM (C-AFM) on CIPS has also shown the formation of both reversible[15,25,57] and irreversible topographical features[58], and such topographical changes were seen during our own PFM and KPFM measurements on CCPS (Figure S13). Raman spectroscopy measurements (plotted in Figure 6c) taken on the edges of the overcycled Device 2 show the clear emergence of peaks related to the S-S stretching modes in amorphous $S_8$[56,59] (red shaded regions). Based on the appearance of these peaks, we can conclude that in the soft-breakdown state measured above, a significant amount of $Cu^+$ is removed from the lattice, after which irreversible restructuring occurs. However, the reversibility of surface features formed in C-AFM indicates that an operation regime exists where no permanent damage occurs. Figure 6c,d show a separate device (Device 5) before and after measuring, for which the resistive switching curves are plotted in Figure S14. In this case, a hard breakdown (shorting) of the device occurred under vacuum, but no visible topographical changes or damage to the electrode are observed in the optical microscope image. Raman spectra measured on the edge of the electrode are plotted in Figure 6d and are clearly free from $S_8$ related peaks. These results highlight that while complex electrochemical reactions occur within CCPS, and may even be intrinsic to the resistive switching, there is a range of operating conditions under which the switching is fully reversible (attributed to the reversibility of the interfacial redox), so that it can be exploited in memory devices without sacrificing endurance.

**Conclusions**

In this paper, we have studied the rate- and voltage-dependence of resistive switching in CCPS and developed a qualitative model to explain the device dynamics, based on ion migration and interfacial electrochemistry. Fitting the readout curves of devices based on CCPS under different switching conditions, we found that both the magnitude of the conductivity as well as the volatility of the low-resistance states can be tuned. By demonstrating resistive switching on CCPS in the paraelectric state, we show that robust resistive switching occurs even in the absence of ferroelectricity. This opens the possibility for integrating devices with different, independently accessible switching mechanisms and different operation timescales as dual-mode devices on the same material platform, as was recently demonstrated for HZO[8]. Since the resistive switching is related to native $Cu^+$ it furthermore bypasses the need for forming steps which are required in that materials system[9,60]. We have directly measured signatures of interfacial redox in Cr/CCPS/Au and Pt/CCPS/Au systems and shown that it is a reversible, diffusion-limited reaction, justifying the observation of filaments in literature[11] and in our data, and offering another route for tuning the conductance volatility. The reversibility of the redox reaction indicates that these electrochemical effects can be harnessed in memory devices without sacrificing device endurance. A better understanding of the ion dynamics and electrochemical behavior of the CCPS material system is an important step towards demonstrating controllable ionic devices for real-time neuromorphic processing. Since ion dynamics are strongly temperature-dependent, further investigations into the temperature dependence of conduction and ionic relaxation are also needed.

We also observed, however, that irreversible electrochemistry can occur, which can be seen through the emergence of additional peaks in the Raman spectrum of the material after cycling. This highlights that careful control over the switching parameters is needed in order to avoid over-extraction of the native $Cu^+$ from the lattice. At the same time, our observation of bipolar redox has important implications for ionic control of ferroelectricity. In CIPS, CCPS or other related materials, where ferroelectricity can be switched 'on' and 'off' by mobile ions[31,32], the redox pathways identified here will further modulate that behavior. Ultimately, the ability to electrically resolve redox in CCPS (even with inert contacts), its influence on ionic switching, and the confirmation of bipolar processes establish a coherent picture of the device physics and can inform future device design and operation.

## Methods

### Sample fabrication

Bottom electrodes (BEs) of 2 nm Ti/10 nm Au were defined via E-beam lithography (EBL), e-beam/thermal evaporation and lift-off on substrates of Si/300 nm thermal $SiO_2$. CIPS and CCPS single crystals from HQ Graphene were exfoliated to produce thinner flakes, identified in an optical microscope. Chosen flakes were stamped onto the BEs using PDMS in an Ar atmosphere. Finally, individual flakes were identified in an optical microscope, for which a second EBL and evaporation step was used to form top electrodes (TEs) composed of 5 nm Cr/50 nm Au unless otherwise stated.

### Micro-Raman spectroscopy

Raman spectroscopy measurements were performed either on devices where stated, or on flakes of CIPS/CCPS exfoliated directly onto BEs covered in blanket 2 nm Ti/10 nm Au deposited via e-beam/thermal evaporation. These measurements were performed in a Renishaw inVia Qontor confocal system with a Nikon 50x long working distance objective and a 633 nm laser source. The power was kept at ca. 0.5 mW which was found to cause minimal heating at all thicknesses studied. Each individual spectrum was averaged over 10 integrations of 1 s.

### Electrical characterization

Electrical characterization was performed in ambient atmosphere with a Keithley 2635A sourcemeter or under a vacuum of 0.12 mbar with a Keithley 4200 Semiconductor Characterization System. In both cases, top and bottom electrodes were contacted with W needles. Schematics of the electrical pulsing schemes applied can be found in Supplementary Figure S3.

**Piezoresponse Force Microscopy (PFM/ Kelvin Probe Force Microscopy (KPFM)**

PFM/KPFM measurements were performed on flakes of CIPS and CCPS exfoliated directly onto substrates of Si/SiO2 covered with thin Au films using a Park NX10 SPM. Both techniques were performed with conductive cantilevers, namely NCS36 Cr/Au ($k = 0.6$-$2$ N/m, $f_R = 60$-$90$ kHz) and CDT-CONTR (diamond-coated tips, $k = 0.5$ N/m, $f_R = 20$ kHz), NCS36 were preferred for KPFM, while CDT-CONTR for PFM. The maps were acquired both in ambient conditions and in Ar atmosphere. KPFM maps were acquired while scanning in non-contact with an AC tip bias $V_{ac} = 1$V @ 17 kHz and with the substrate grounded. PFM maps were acquired w/wo DC sample bias to write/read the ferroelectric domains and with $V_{ac} = 1$-$1.5$ V. To enhance the PFM signals the maps in contact were acquired in-resonance conditions (97-104 kHz).

**Atomic Force Microscopy**

Atomic Force Microscopy (AFM) images were recorded directly on devices after electrical characterization on a JPK Bruker system with APPNano ACT-50 n-type Si tip. They were recorded in tapping mode, at room temperature in an ambient atmosphere. AFM images were post-processed in Gwyddion.

ASSOCIATED CONTENT

The following files are available free of charge.

Supplementary materials (PDF): Details of devices (flake thicknesses, AFM measurements, device areas); Piezoresponse Force Microscopy on CIPS and CCPS; electrical measurement schemes; resistive switching measurements (frequency dependence in positive/negative polarity, voltage dependence); time constants extracted from retention measurements; measurements of redox and filamentary signals in devices with Cr or Pt top electrodes; diffusion coefficient calculations (from data and literature); Kelvin Probe Force Microscopy measurements on CCPS.

AUTHOR INFORMATION

**Corresponding Author**

*Corresponding author: Suzanne Lancaster. Email: *s.lancaster@nanogune.eu*

**Author Contributions**

The manuscript was written through contributions of all authors. All authors have given approval to the final version of the manuscript.

**Funding Sources**

S.L and T.A. acknowledge funding from the European Union under Marie Skłodowska-Curie grant agreements no. 101151842 (ESTRELLA) and no. 101107842 (ACCESS), respectively. The authors acknowledge funding from MICIU/AEI/10.13039/501100011033 (grant CEX2020-001038-M); from MICIU/AEI and ERDF/EU (projects PID2024-155708OB-I00, PID2024-157558NB-C21, PID2024-157558NB-C22); from the Italian Space Agency (ASI) through "SPACE IT UP!" (contract N. 2024-5-E.0); and from the European Union's Horizon 2020 research and innovation program under the Marie Skłodowska-Curie grant agreement No. 955671. L.O.V. acknowledges funding from MICIU/AEI and ESF+ (fellowship PRE2022-104385). B.M.-G. and M.G. acknowledge funding from MICIU/AEI and European Union NextGenerationEU/PRTR (grants RYC2021-034836-I and RYC2021-031705-I, respectively).

ACKNOWLEDGMENT

The authors would like to thank Dr. Shayan Edalatmanesh and Dr. Ane Etxebarria Dueñas for fruitful discussions.

REFERENCES

(1) Indiveri, G.; Linares-Barranco, B.; Hamilton, T. J.; van Schaik, A.; Etienne-Cummings, R.; Delbruck, T.; Liu, S.-C.; Dudek, P.; Häfliger, P.; Renaud, S.; Schemmel, J.; Cauwenberghs, G.; Arthur, J.; Hynna, K.; Folowosele, F.; Saïghi, S.; Serrano-Gotarredona, T.; Wijekoon, J.; Wang, Y.;

Boahen, K. Neuromorphic Silicon Neuron Circuits. *Front. Neurosci.* **2011**, *5*. https://doi.org/10.3389/fnins.2011.00073.
(2) Ielmini, D.; Ambrogio, S. Emerging Neuromorphic Devices. *Nanotechnology* **2019**, *31* (9), 092001. https://doi.org/10.1088/1361-6528/ab554b.
(3) Covi, E.; Lancaster, S.; Slesazeck, S.; Deshpande, V.; Mikolajick, T.; Dubourdieu, C. Challenges and Perspectives for Energy-Efficient Brain-Inspired Edge Computing Applications (Invited Paper). In *2022 IEEE International Conference on Flexible and Printable Sensors and Systems (FLEPS)*; 2022; pp 1–4. https://doi.org/10.1109/FLEPS53764.2022.9781597.
(4) Pati, S. P.; Yajima, T. Review of Solid-State Proton Devices for Neuromorphic Information Processing. *Jpn. J. Appl. Phys.* **2024**, *63* (3), 030801. https://doi.org/10.35848/1347-4065/ad297b.
(5) Pati, S. P.; Geng, Y.; Hamasuna, S.; Fujiwara, K.; Iizuka, T.; Inoue, H.; Inoue, I.; Yajima, T. Real-Time Information Processing via Volatile Resistance Change in Scalable Protonic Devices. *Commun. Mater.* **2024**, *5* (1), 177. https://doi.org/10.1038/s43246-024-00621-1.
(6) Halter, M.; Bégon-Lours, L.; Sousa, M.; Popoff, Y.; Drechsler, U.; Bragaglia, V.; Offrein, B. J. A Multi-Timescale Synaptic Weight Based on Ferroelectric Hafnium Zirconium Oxide. *Commun. Mater.* **2023**, *4* (1), 14. https://doi.org/10.1038/s43246-023-00342-x.
(7) Chen, Y.; Han, B.; Gobbi, M.; Hou, L.; Samorì, P. Responsive Molecules for Organic Neuromorphic Devices: Harnessing Memory Diversification. *Adv. Mater.* **2025**, *37* (19), 2418281. https://doi.org/10.1002/adma.202418281.
(8) Martemucci, M.; Rummens, F.; Malot, Y.; Hirtzlin, T.; Guille, O.; Martin, S.; Carabasse, C.; Vincent, A. F.; Saïghi, S.; Grenouillet, L.; Querlioz, D.; Vianello, E. A Ferroelectric–Memristor Memory for Both Training and Inference. *Nat. Electron.* **2025**, *8* (10), 921–933. https://doi.org/10.1038/s41928-025-01454-7.
(9) Knabe, J.; Berg, F.; Thorben Goβ, K.; Boettger, U.; Dittmann, R. Dual-Mode Operation of Epitaxial Hf0.5Zr0.5O2: Ferroelectric and Filamentary-Type Resistive Switching. *Phys. Status Solidi A* **2024**, *221* (22), 2300409. https://doi.org/10.1002/pssa.202300409.
(10) Jiang, X.; Zhang, X.; Deng, Z.; Deng, J.; Wang, X.; Wang, X.; Yang, W. Dual-Role Ion Dynamics in Ferroionic CuInP2S6: Revealing the Transition from Ferroelectric to Ionic Switching Mechanisms. *Nat. Commun.* **2024**, *15* (1), 10822. https://doi.org/10.1038/s41467-024-55160-7.
(11) Zhou, J.; Chen, A.; Zhang, Y.; Pu, D.; Qiao, B.; Hu, J.; Li, H.; Zhong, S.; Zhao, R.; Xue, F.; Xu, Y.; Loh, K. P.; Wang, H.; Yu, B. 2D Ferroionics: Conductive Switching Mechanisms and Transition Boundaries in Van Der Waals Layered Material CuInP2S6. *Adv. Mater.* **2023**, *35* (38), 2302419. https://doi.org/10.1002/adma.202302419.
(12) Ma, R.-R.; Xu, D.-D.; Zhong, Q.-L.; Zhong, C.-R.; Huang, R.; Xiang, P.-H.; Zhong, N.; Duan, C.-G. Nanoscale Mapping of Cu-Ion Transport in van Der Waals Layered CuCrP2S6. *Adv. Mater. Interfaces* **2022**, *9* (4), 2101769. https://doi.org/10.1002/admi.202101769.
(13) Jiang, X.; Zhang, X.; Deng, Z.; Deng, J.; Wang, X.; Wang, X.; Yang, W. Dual-Role Ion Dynamics in Ferroionic CuInP2S6: Revealing the Transition from Ferroelectric to Ionic Switching Mechanisms. *Nat. Commun.* **2024**, *15* (1), 10822. https://doi.org/10.1038/s41467-024-55160-7.
(14) Zhang, D.; Luo, Z.-D.; Yao, Y.; Schoenherr, P.; Sha, C.; Pan, Y.; Sharma, P.; Alexe, M.; Seidel, J. Anisotropic Ion Migration and Electronic Conduction in van Der Waals Ferroelectric CuInP2S6. *Nano Lett.* **2021**, *21* (2), 995–1002. https://doi.org/10.1021/acs.nanolett.0c04023.
(15) Balke, N.; Neumayer, S. M.; Brehm, J. A.; Susner, M. A.; Rodriguez, B. J.; Jesse, S.; Kalinin, S. V.; Pantelides, S. T.; McGuire, M. A.; Maksymovych, P. Locally Controlled Cu-Ion Transport in Layered Ferroelectric CuInP2S6. *ACS Appl. Mater. Interfaces* **2018**, *10* (32), 27188–27194. https://doi.org/10.1021/acsami.8b08079.
(16) Guo, C.; Zhu, J.; Liang, X.; Wen, C.; Xie, J.; Gu, C.; Hu, W. Atomic-Level Direct Imaging for Cu(I) Multiple Occupations and Migration in 2D Ferroelectric CuInP2S6. *Nat. Commun.* **2024**, *15* (1), 10152. https://doi.org/10.1038/s41467-024-54229-7.
(17) Wu, X.; Han, S.; Zhang, Z.; Wang, Y.; Li, J.; Zhang, Z.; Wang, D.; Peng, X.; She, Y.; Gao, X.; Zhou, C.; Wang, G.; Zhang, Z. Rejuvenation of Mechanical Fatigue Resistance in 2D Ferroelectric

CuInP2S6 by Reversing Ionic Motion. *Small* **2026**, *22* (30), e73338. https://doi.org/10.1002/smll.73338.
(18) Maisonneuve, V.; Cajipe, V. B.; Simon, A.; Von Der Muhll, R.; Ravez, J. Ferrielectric Ordering in Lamellar CuCrP2S6 *Phys. Rev. B* **1997**, *56* (17), 10860–10868. https://doi.org/10.1103/PhysRevB.56.10860.
(19) Cajipea, V. B.; Ravez, J.; Maisonneuve, V.; Simon, A.; Payen, C.; Von Der Muhll, R.; Fischer, J. E. Copper Ordering in Lamellar CuMP2S6 (M= Cr, In): Transition to an Antiferroelectric or Ferroelectric Phase. *Ferroelectrics* **1996**, *185* (1), 135–138. https://doi.org/10.1080/00150199608210497.
(20) Silva, J. P. B.; Alcala, R.; Avci, U. E.; Barrett, N.; Bégon-Lours, L.; Borg, M.; Byun, S.; Chang, S.-C.; Cheong, S.-W.; Choe, D.-H.; Coignus, J.; Deshpande, V.; Dimoulas, A.; Dubourdieu, C.; Fina, I.; Funakubo, H.; Grenouillet, L.; Gruverman, A.; Heo, J.; Hoffmann, M.; Hsain, H. A.; Huang, F.-T.; Hwang, C. S.; Íñiguez, J.; Jones, J. L.; Karpov, I. V.; Kersch, A.; Kwon, T.; Lancaster, S.; Lederer, M.; Lee, Y.; Lomenzo, P. D.; Martin, L. W.; Martin, S.; Migita, S.; Mikolajick, T.; Noheda, B.; Park, M. H.; Rabe, K. M.; Salahuddin, S.; Sánchez, F.; Seidel, K.; Shimizu, T.; Shiraishi, T.; Slesazeck, S.; Toriumi, A.; Uchida, H.; Vilquin, B.; Xu, X.; Ye, K. H.; Schroeder, U. Roadmap on Ferroelectric Hafnia- and Zirconia-Based Materials and Devices. *APL Mater.* **2023**, *11* (8), 089201. https://doi.org/10.1063/5.0148068.
(21) Fichtner, S.; Wolff, N.; Lofink, F.; Kienle, L.; Wagner, B. AlScN: A III-V Semiconductor Based Ferroelectric. *J. Appl. Phys.* **2019**, *125* (11), 114103. https://doi.org/10.1063/1.5084945.
(22) Tu, B. Q.; Ahmed, T.; Avedissian, G.; Lancaster, S.; Sharma, M.; Watanabe, K.; Taniguchi, T.; Casanova, F.; Gobbi, M.; Hueso, L. E. Ferroelectric Hysteresis in Singly Aligned Graphene-hBN Moiré Superlattices. *Small* **2025**, *21* (46), e08416. https://doi.org/10.1002/smll.202508416.
(23) Ding, W.; Zhu, J.; Wang, Z.; Gao, Y.; Xiao, D.; Gu, Y.; Zhang, Z.; Zhu, W. Prediction of Intrinsic Two-Dimensional Ferroelectrics in In2Se3 and Other III2-VI3 van Der Waals Materials. *Nat. Commun.* **2017**, *8* (1), 14956. https://doi.org/10.1038/ncomms14956.
(24) Liu, F.; You, L.; Seyler, K. L.; Li, X.; Yu, P.; Lin, J.; Wang, X.; Zhou, J.; Wang, H.; He, H.; Pantelides, S. T.; Zhou, W.; Sharma, P.; Xu, X.; Ajayan, P. M.; Wang, J.; Liu, Z. Room-Temperature Ferroelectricity in CuInP2S6 Ultrathin Flakes. *Nat. Commun.* **2016**, *7* (1), 12357. https://doi.org/10.1038/ncomms12357.
(25) Belianinov, A.; He, Q.; Dziaugys, A.; Maksymovych, P.; Eliseev, E.; Borisevich, A.; Morozovska, A.; Banys, J.; Vysochanskii, Y.; Kalinin, S. V. CuInP2S6 Room Temperature Layered Ferroelectric. *Nano Lett.* **2015**, *15* (6), 3808–3814. https://doi.org/10.1021/acs.nanolett.5b00491.
(26) Li, B.; Li, S.; Wang, H.; Chen, L.; Liu, L.; Feng, X.; Li, Y.; Chen, J.; Gong, X.; Ang, K.-W. An Electronic Synapse Based on 2D Ferroelectric CuInP2S6. *Adv. Electron. Mater.* **2020**, *6* (12), 2000760. https://doi.org/10.1002/aelm.202000760.
(27) Io, W. F.; Pang, S.-Y.; Wong, L. W.; Zhao, Y.; Ding, R.; Mao, J.; Zhao, Y.; Guo, F.; Yuan, S.; Zhao, J.; Yi, J.; Hao, J. Direct Observation of Intrinsic Room-Temperature Ferroelectricity in 2D Layered CuCrP2S6. *Nat. Commun.* **2023**, *14* (1), 7304. https://doi.org/10.1038/s41467-023-43097-2.
(28) Cho, K.; Lee, S.; Kalaivanan, R.; Sankar, R.; Choi, K.-Y.; Park, S. Tunable Ferroelectricity in Van Der Waals Layered Antiferroelectric CuCrP2S6. *Adv. Funct. Mater.* **2022**, *32* (36), 2204214. https://doi.org/10.1002/adfm.202204214.
(29) Susner, M. A.; Rao, R.; Pelton, A. T.; McLeod, M. V.; Maruyama, B. Temperature-Dependent Raman Scattering and x-Ray Diffraction Study of Phase Transitions in Layered Multiferroic CuCrP2S6. *Phys. Rev. Mater.* **2020**, *4* (10), 104003. https://doi.org/10.1103/PhysRevMaterials.4.104003.
(30) Neumayer, S.; Qiao, H.; Balke, N. Competing Polar Phases in 2D Ferroelectric Transition Metal Thio- and Selenophosphates. *Appl. Phys. Lett.* **2025**, *126* (12), 120501. https://doi.org/10.1063/5.0253879.

(31) Neumayer, S. M.; Si, M.; Li, J.; Liao, P.-Y.; Tao, L.; O'Hara, A.; Pantelides, S. T.; Ye, P. D.; Maksymovych, P.; Balke, N. Ionic Control over Ferroelectricity in 2D Layered van Der Waals Capacitors. *ACS Appl. Mater. Interfaces* **2022**, *14* (2), 3018–3026. https://doi.org/10.1021/acsami.1c18683.
(32) Sun, F.; Xu, H.; Ju, Q.; Hong, W.; Cai, Q.; Sun, Z.; Liu, W. Field-Induced Interlayer Ion Migration and Electronic Coupling Unlock Ferroelectricity in Centrosymmetric AgInP2Se6 Crystals. *J. Am. Chem. Soc.* **2025**, *147* (30), 26804–26812. https://doi.org/10.1021/jacs.5c07965.
(33) Wang, Z.; Joshi, S.; Savel'ev, S. E.; Jiang, H.; Midya, R.; Lin, P.; Hu, M.; Ge, N.; Strachan, J. P.; Li, Z.; Wu, Q.; Barnell, M.; Li, G.-L.; Xin, H. L.; Williams, R. S.; Xia, Q.; Yang, J. J. Memristors with Diffusive Dynamics as Synaptic Emulators for Neuromorphic Computing. *Nat. Mater.* **2017**, *16* (1), 101–108. https://doi.org/10.1038/nmat4756.
(34) Qin, L.; Yu, Y.; Fang, C.; Liu, Y.; Zhu, K.; Ouyang, D.; Liu, S.; Song, B.; Zhou, R.; Lanza, M.; Hu, W.; Wu, J.; Li, Y.; Zhai, T. Intrinsic Ion Migration-Induced Susceptible Two-Dimensional Phase-Transition Memristor with Ultralow Power Consumption. *Sci. Bull.* **2025**, *70* (13), 2116–2124. https://doi.org/10.1016/j.scib.2025.03.051.
(35) Ma, Y.; Yan, Y.; Luo, L.; Pazos, S.; Zhang, C.; Lv, X.; Chen, M.; Liu, C.; Wang, Y.; Chen, A.; Li, Y.; Zheng, D.; Lin, R.; Algaidi, H.; Sun, M.; Liu, J. Z.; Tu, S.; Alshareef, H. N.; Gong, C.; Lanza, M.; Xue, F.; Zhang, X. High-Performance van Der Waals Antiferroelectric CuCrP2S6-Based Memristors. *Nat. Commun.* **2023**, *14* (1), 7891. https://doi.org/10.1038/s41467-023-43628-x.
(36) Liu, P.; Li, Y.; Hou, D.; Zhu, H.; Luo, H.; Zhou, S.; Wei, L.; Niu, W.; Sheng, Z.; Mao, W.; Pu, Y. Switchable Diode Effect in 2D van Der Waals Ferroelectric CuCrP2S6. *Appl. Phys. Lett.* **2024**, *124* (9). https://doi.org/10.1063/5.0191188.
(37) Watson, A. J.; Lu, W.; Guimarães, M. H. D.; Stöhr, M. Transfer of Large-Scale Two-Dimensional Semiconductors: Challenges and Developments. *2D Mater.* **2021**, *8* (3), 032001. https://doi.org/10.1088/2053-1583/abf234.
(38) Hong, M.; Dai, L.; Hu, H.; Li, C. Structural, Ferroelectric, and Electronic Transitions in the van Der Waals Multiferroic Material CuCrP2S6 under High Temperature and High Pressure. *Phys. Rev. B* **2024**, *110* (14), 144103. https://doi.org/10.1103/PhysRevB.110.144103.
(39) Lancaster, S.; Remillieux, M.; Engl, M.; Havel, V.; Silva, C.; Wang, X.; Mikolajick, T.; Slesazeck, S. Weight Update in Ferroelectric Memristors with Identical and Nonidentical Pulses. *ACS Appl. Mater. Interfaces* **2024**, *16* (38), 51109–51117. https://doi.org/10.1021/acsami.4c10338.
(40) Kohlrausch, R. Theorie Des Elektrischen Rückstandes in Der Leidener Flasche. *Ann. Phys.* **1854**, *167* (2), 179–214. https://doi.org/10.1002/andp.18541670203.
(41) Williams, G.; Watts, D. C. Non-Symmetrical Dielectric Relaxation Behaviour Arising from a Simple Empirical Decay Function. *Trans. Faraday Soc.* **1970**, *66* (0), 80–85. https://doi.org/10.1039/TF9706600080.
(42) Apitz, D.; Johansen, P. M. Limitations of the Stretched Exponential Function for Describing Dynamics in Disordered Solid Materials. *J. Appl. Phys.* **2005**, *97* (6), 063507. https://doi.org/10.1063/1.1852069.
(43) León, C.; Lucía, M. L.; Santamaría, J.; Sánchez-Quesada, F. Universal Scaling of the Conductivity Relaxation in Crystalline Ionic Conductors. *Phys. Rev. B* **1998**, *57* (1), 41–44. https://doi.org/10.1103/PhysRevB.57.41.
(44) Zhu, H.; Li, J.; Chen, Q.; Tang, W.; Fan, X.; Li, F.; Li, L. Highly Tunable Lateral Homojunction Formed in Two-Dimensional Layered CuInP2S6 via In-Plane Ionic Migration. *ACS Nano* **2023**, *17* (2), 1239–1246. https://doi.org/10.1021/acsnano.2c09280.
(45) Zhong, Z.; Wu, S.; Li, X.; Wang, Z.; Yang, Q.; Huang, B.; Chen, Y.; Wang, X.; Lin, T.; Shen, H.; Meng, X.; Wang, M.; Shi, W.; Wang, J.; Chu, J.; Huang, H. Robust Threshold-Switching Behavior Assisted by Cu Migration in a Ferroionic CuInP2S6 Heterostructure. *ACS Nano* **2023**, *17* (13), 12563–12572. https://doi.org/10.1021/acsnano.3c02406.

(46) Zhu, H.; Li, J.; Chen, Q.; Tang, W.; Fan, X.; Li, F.; Li, L. Highly Tunable Lateral Homojunction Formed in Two-Dimensional Layered CuInP2S6 via In-Plane Ionic Migration. *ACS Nano* **2023**, *17* (2), 1239–1246. https://doi.org/10.1021/acsnano.2c09280.
(47) Yamada, H.; Yoshii, K.; Asahi, M.; Chiku, M.; Kitazumi, Y. Cyclic Voltammetry Part 1: Fundamentals. *Electrochemistry* **2022**, *90* (10), 102005–102005. https://doi.org/10.5796/electrochemistry.22-66082.
(48) Elgrishi, N.; Rountree, K. J.; McCarthy, B. D.; Rountree, E. S.; Eisenhart, T. T.; Dempsey, J. L. A Practical Beginner's Guide to Cyclic Voltammetry. *J. Chem. Educ.* **2018**, *95* (2), 197–206. https://doi.org/10.1021/acs.jchemed.7b00361.
(49) Maisonneuve, V.; Reau, J. M.; Dong, M.; Cajipe, V. B.; Payen, C.; Ravez, J. Ionic Conductivity in Ferroic CuInP2S6 and CuCrP2S6. *Ferroelectrics* **1997**, *196* (1), 257–260. https://doi.org/10.1080/00150199708224175.
(50) Dziaugys, A.; Banys, J.; Macutkevic, J.; Vysochanskii, Y. Anisotropy Effects in Thick Layered CuInP2S6 and CuInP2Se6 Crystals. *Phase Transit.* **2013**, *86* (9), 878–885. https://doi.org/10.1080/01411594.2012.745533.
(51) Zhou, Z.; Zhang, J.-J.; Turner, G. F.; Moggach, S. A.; Lekina, Y.; Morris, S.; Wang, S.; Hu, Y.; Li, Q.; Xue, J.; Feng, Z.; Yan, Q.; Weng, Y.; Xu, B.; Fang, Y.; Shen, Z. X.; Fang, L.; Dong, S.; You, L. Sliding-Mediated Ferroelectric Phase Transition in CuInP2S6 under Pressure. *Appl. Phys. Rev.* **2024**, *11* (1), 011414. https://doi.org/10.1063/5.0177451.
(52) Tappertzhofen, S.; Valov, I.; Tsuruoka, T.; Hasegawa, T.; Waser, R.; Aono, M. Generic Relevance of Counter Charges for Cation-Based Nanoscale Resistive Switching Memories. *ACS Nano* **2013**, *7* (7), 6396–6402. https://doi.org/10.1021/nn4026614.
(53) Lübben, M.; Menzel, S.; Park, S. G.; Yang, M.; Waser, R.; Valov, I. SET Kinetics of Electrochemical Metallization Cells: Influence of Counter-Electrodes in SiO2/Ag Based Systems. *Nanotechnology* **2017**, *28* (13), 135205. https://doi.org/10.1088/1361-6528/aa5e59.
(54) O'Hara, A.; Tao, L.; Neumayer, S. M.; Maksymovych, P.; Balke, N.; Pantelides, S. T. Effects of Thin Metal Contacts on Few-Layer van Der Waals Ferrielectric CuInP2S6. *J. Appl. Phys.* **2022**, *132* (11), 114102. https://doi.org/10.1063/5.0096704.
(55) Schindler, C.; Thermadam, S. C. P.; Waser, R.; Kozicki, M. N. Bipolar and Unipolar Resistive Switching in Cu-Doped \hboxSiO_2. *IEEE Trans. Electron Devices* **2007**, *54* (10), 2762–2768. https://doi.org/10.1109/TED.2007.904402.
(56) Zhou, Z.; Wang, S.; Zhou, Z.; Hu, Y.; Li, Q.; Xue, J.; Feng, Z.; Yan, Q.; Luo, Z.; Weng, Y.; Tang, R.; Su, X.; Zheng, F.; Okamoto, K.; Funakubo, H.; Kang, L.; Fang, L.; You, L. Unconventional Polarization Fatigue in van Der Waals Layered Ferroelectric Ionic Conductor CuInP2S6. *Nat. Commun.* **2023**, *14* (1), 8254. https://doi.org/10.1038/s41467-023-44132-y.
(57) Jiang, X.; Wang, X.; Wang, X.; Zhang, X.; Niu, R.; Deng, J.; Xu, S.; Lun, Y.; Liu, Y.; Xia, T.; Lu, J.; Hong, J. Manipulation of Current Rectification in van Der Waals Ferroionic CuInP2S6. *Nat. Commun.* **2022**, *13* (1), 574. https://doi.org/10.1038/s41467-022-28235-6.
(58) Ming, W.; Huang, B.; Zheng, S.; Bai, Y.; Wang, J.; Wang, J.; Li, J. Flexoelectric Engineering of van Der Waals Ferroelectric CuInP2S6. *Sci. Adv.* **2022**, *8* (33), eabq1232. https://doi.org/10.1126/sciadv.abq1232.
(59) Crapanzano, L. Polymorphism of Sulfur: Structural and Dynamical Aspects. Theses, Université Joseph-Fourier - Grenoble I, 2006. https://theses.hal.science/tel-00204149 (accessed 2025-12-19).
(60) Januel, T.; Billoint, O.; Grenouillet, L.; Martin, S.; Portal, J. M.; Vianello, E. Dual-Mode 16kb Memory: Transforming a Ferroelectric Capacitor Bitcell into Resistive Filamentary Memory. In *2025 IEEE International Memory Workshop (IMW)*; 2025; pp 1–4. https://doi.org/10.1109/IMW61990.2025.11026985.

# Supporting information: Rate-Programmable Ionic–Redox Switching with Tunable Volatility in $CuCrP_2S_6$

**Suzanne Lancaster[1*], Francesco Calavalle[2], Mayank Sharma[1,3], Lucía Olano-Vegas[1,3], Garen Avedissian[1,4], Tanweer Ahmed[1], Marco Gobbi[5,6], Beatriz Martín-García[1,5,6], Beatrice Fraboni[2,7], Fèlix Casanova[1,6], Luis E. Hueso[1,6]**

*1 – CIC NanoGUNE BRTA, 20018 Donostia-San Sebastián, Basque Country, Spain*
*2 - Department of Physics and Astronomy, University of Bologna, Viale Berti Pichat 6/2, Bologna, 40127 Italy*
*3 - Departamento de Polímeros y Materiales Avanzados: Física, Química y Tecnología, University of the Basque Country (UPV/EHU), Donostia-San Sebastian, 20018, Spain*
*4 - Donostia International Physics Center, DIPC, 20018 Donostia-San Sebastian, Basque Country, Spain*
*5 - Materials Physics Center CSIC-UPV/EHU, 20018 Donostia-San Sebastián, Spain*
*6 - IKERBASQUE, Basque Foundation for Science, 48009 Bilbao, Spain*
*7 - INFN, Sezione di Bologna, Viale Berti Pichat 6/2, Bologna, 40127 Italy*
**Corresponding author. Email: s.lancaster@nanogune.eu*

## Table S1

*Table S1: Flake thicknesses and device area of all devices discussed in this paper*

| Device # | Material | Thickness [nm] | # Layers (approx.) | Area [μm$^2$] | Data presented |
|---|---|---|---|---|---|
| 1 | Cr/$CuCrP_2S_6$ (CCPS)/Au | 33 | 51 | 14 | Resistive switching (main) |
| 2 | Cr/CCPS/Au | 17 | 26 | 14 | Redox, filamentary (main) |
| 3 | Cr/CCPS/Au | 154 | 237 | 41 | Redox, Raman (main) |
| 4 | Cr/CCPS/Au | 73 | 112 | 21 | Filamentary (supporting) |
| 5 | Pt/CCPS/Au | 150 | 231 | 31 | Redox (supporting) |
| 6 | Cr/CCPS/Au | 13 | 20 | 13 | Raman (main), resistive switching (supporting) |
| 7 | Cr/$CuInP_2S_6$ (CIPS)/Au | | | 51 | Raman, resistive switching, filamentary (supporting) |
| 8 | Cr/CCPS/Au | 160 | 246 | 12.5 | Resistive switching (supporting) |
| 9 | Cr/CCPS/Au | | | 43 | Retention, time constant extraction (supporting) |
| 10 | Cr/CCPS/Au | | | 69 | Retention, time constant extraction (supporting) |

**Figure S1: Atomic Force Microscopy (AFM) images and thickness extraction**

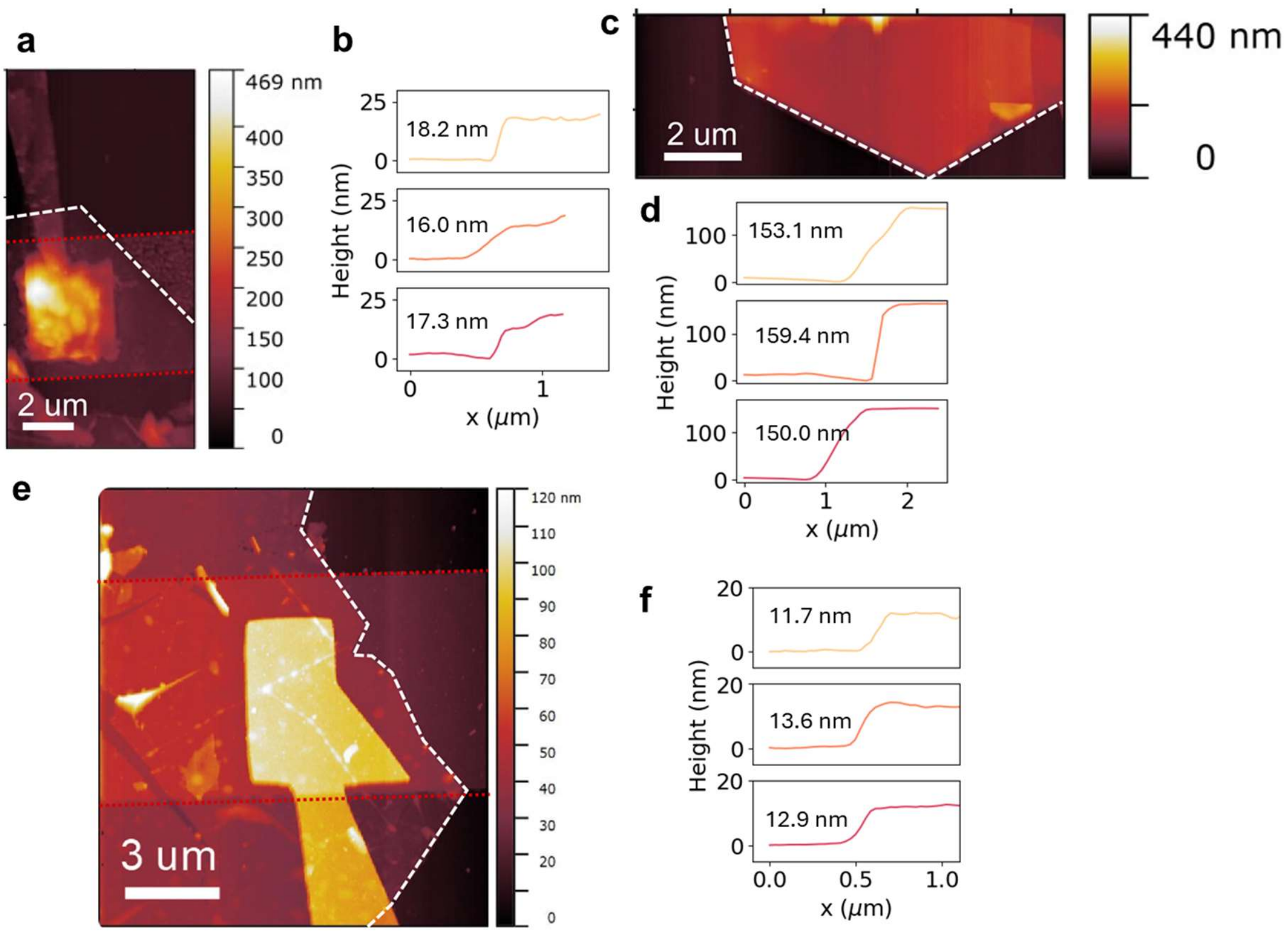

*Figure S1: AFM maps and thickness extraction of devices shown in the main text of this work; (a) AFM map and (b) extracted linescans of Device 2; (c, d) Device 3 and (e, f) Device 6. Red lines indicate the outline of the bottom electrodes, while white dashed lines indicate the outline of CCPS. Note that the AFM images were taken after measurements, so that Devices 3 and 5 had suffered some damage (See main text) and the thickness was extracted away from the electrode.*

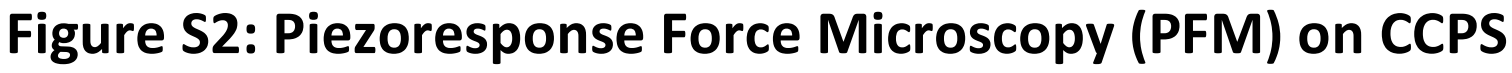

# Figure S2: Piezoresponse Force Microscopy (PFM) on CCPS

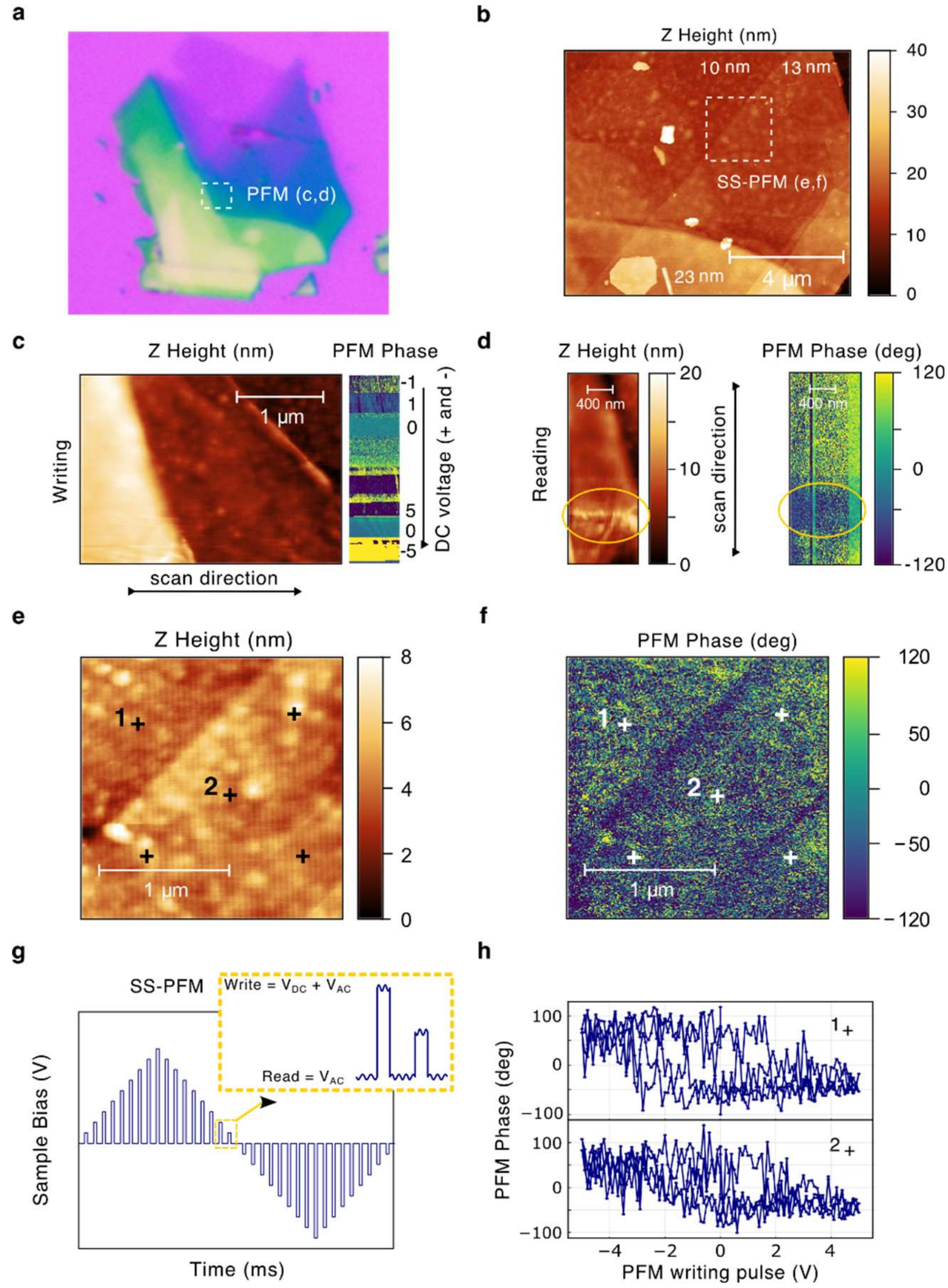

*Figure S2: Representative PFM map and Switching Spectroscopy PFM (SS-PFM) on a CCPS flake. (a) Optical microscope image indicating the area of the PFM maps in (c)-(d); (b) Topographic image of the bottom left area of the flake showing thicknesses and the region where SS-PFM was performed. c) Z-height map of the sample surface acquired while applying the sequence of DC sample bias indicated on the right, the EFM phase during the writing reflect the electrostatic forces between tip and sample surface. (d) Z-height and EFM phase of the same area scanned in (c) but without any DC bias applied (reading scan), the only clear features in the phase map can be correlated with physical damages on the sample surface. (e)-(f) Z-height and EFM phase of the area indicated in (b) at 0 DC bias applied, again no ferroelectric domain is detected and the only clear features in the phase map can be correlated with steps on the sample surface. Markers indicate points at which SS-PFM was performed. (g) Waveform used for SS-PFM. The 'Write time' ($V_{DC}$ ON) for this experiment was 10 ms, while the 'Read' time ($V_{DC}$ OFF) was 20 ms, both parameters were scanned up to 100 ms but no hysteresis loops were detected. The $V_{AC}$ frequency was 97-104 kHz, close to tip resonance. (h) SS-PFM phase spectra at two points marked in (e).*

*There is a slight hysteresis with voltage, with a phase separation of < 180º and a large noise. This hysteresis is likely electrostatic in nature, rather than ferroelectric.*

**Proving the paraelectricity of CCPS:** Figures S2d and f clearly show the absence of oppositely polarised domains after writing, similar to if no contrast is seen in "box-in-box" PFM measurements. On the other hand, on CIPS, regions of oppositely polarised domains with a 180º phase difference are observed when scanning the sample at 0 V external bias (Figure S13b) which is a clear signature of ferroelectricity. In both cases, box-in-box PFM was not performed as the high voltages needed for switching combined with long scanning times lead to damage on the sample, (see e.g. Figure S14e) whereas low-voltage bias scanning clearly shows the absence (CCPS) or presence (CIPS) of ferroelectric domains without leading to ion migration.

**Figure S3: Electrical measurement schemes**

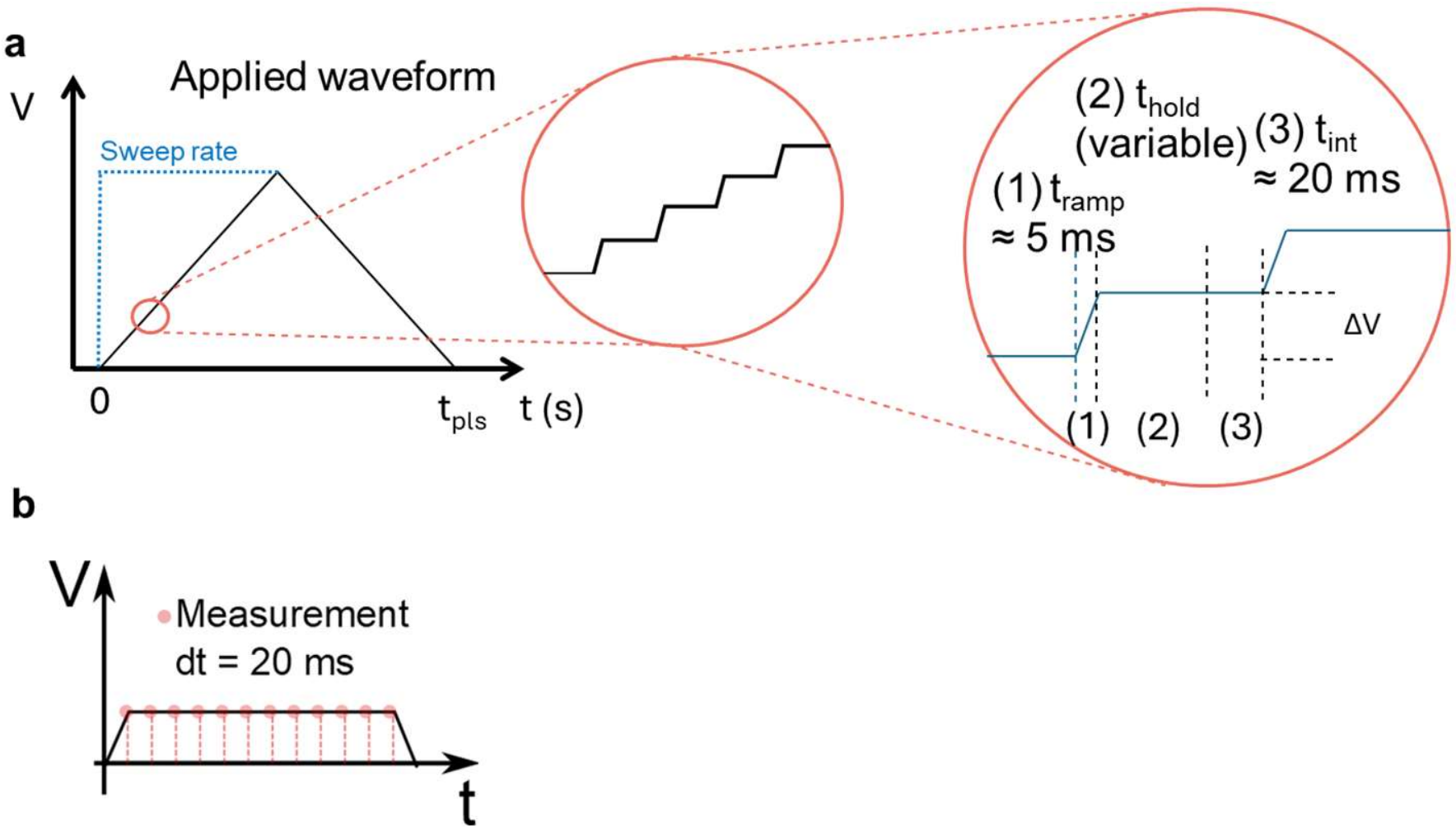

*Figure S3: Schematics of electrical measurement schemes. (a) 'Staircase sweep' used for triangular pulses (with zoom-in on the exact timings used on each step). (b) Measurement scheme used on the 'Square' readout pulses for measuring current in resistive switching regime.*

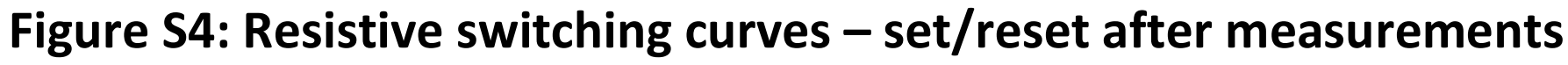

## Figure S4: Resistive switching curves – set/reset after measurements

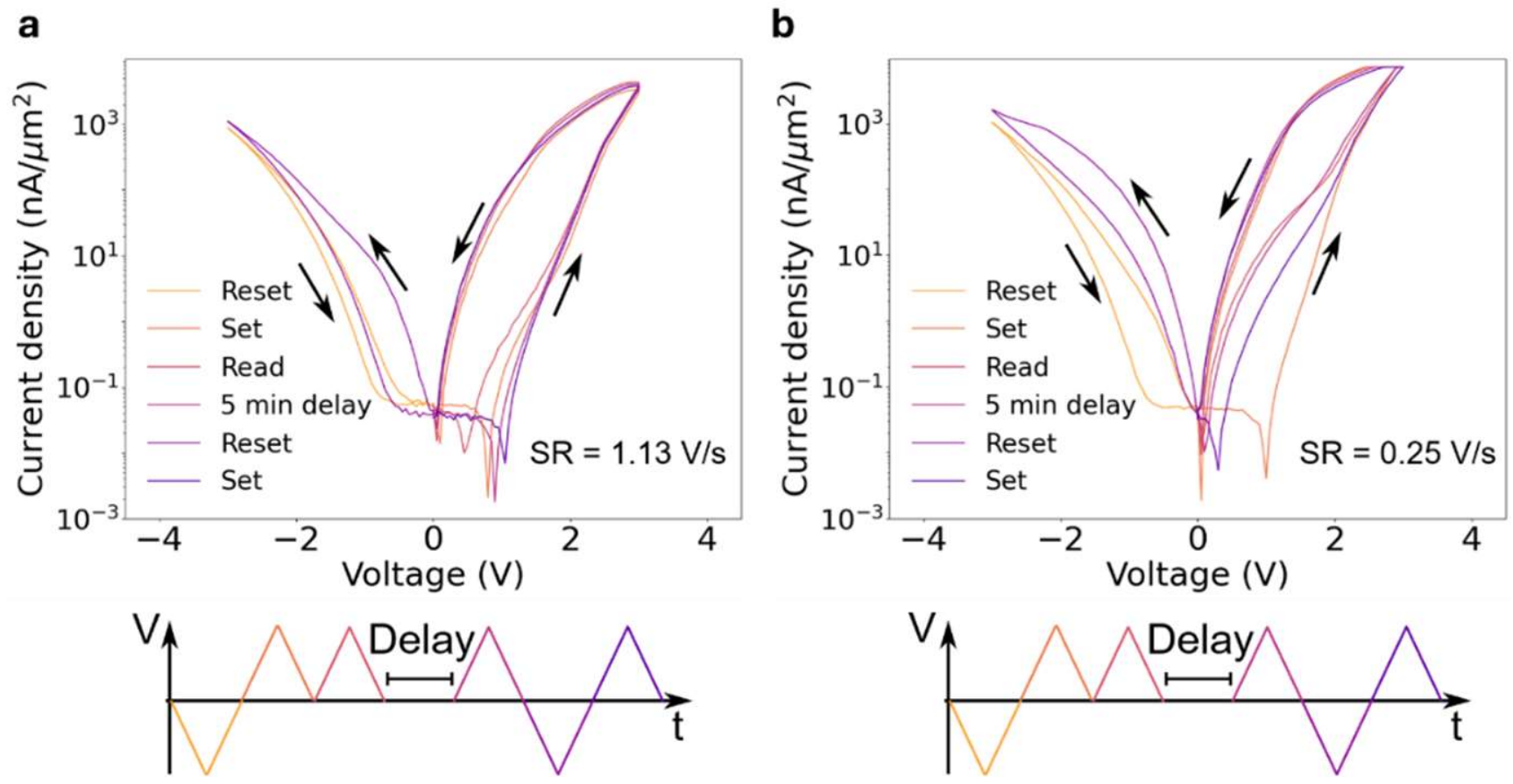

*Figure S4: Set/reset curves measured after the pulse trains shown in Figure 2. The devices were measured at a sweep rate of (a) 1.1.3 V/s and (b) 0.25 V/s. In the case of the faster sweep rate (a), the final reset shows a larger hysteresis due to the temporarily modified device state, but the 'Reset' successfully returns the device to the initial state (as seen by the nearly overlapping traces of the initial 'Set' – Orange, and final 'Set' – purple). In contrast, with a slower sweep rate (b) the initial and final 'Reset' are very different, and it is clear that the device cannot be fully 'Reset' with a single pulse. This can be seen in the negative polarity – on the return sweep of the final 'Reset' (purple), the conductivity is higher than on the initial 'Reset' (yellow). A subsequent 'Set' pulse (purple) also shows a higher conductivity than the initial 'Set' pulse (orange) in positive polarity. These results indicate a more remanent modification to the device state at slower sweep rates.*

## Figure S5: Resistive switching – additional sweep rates, delays in negative polarity

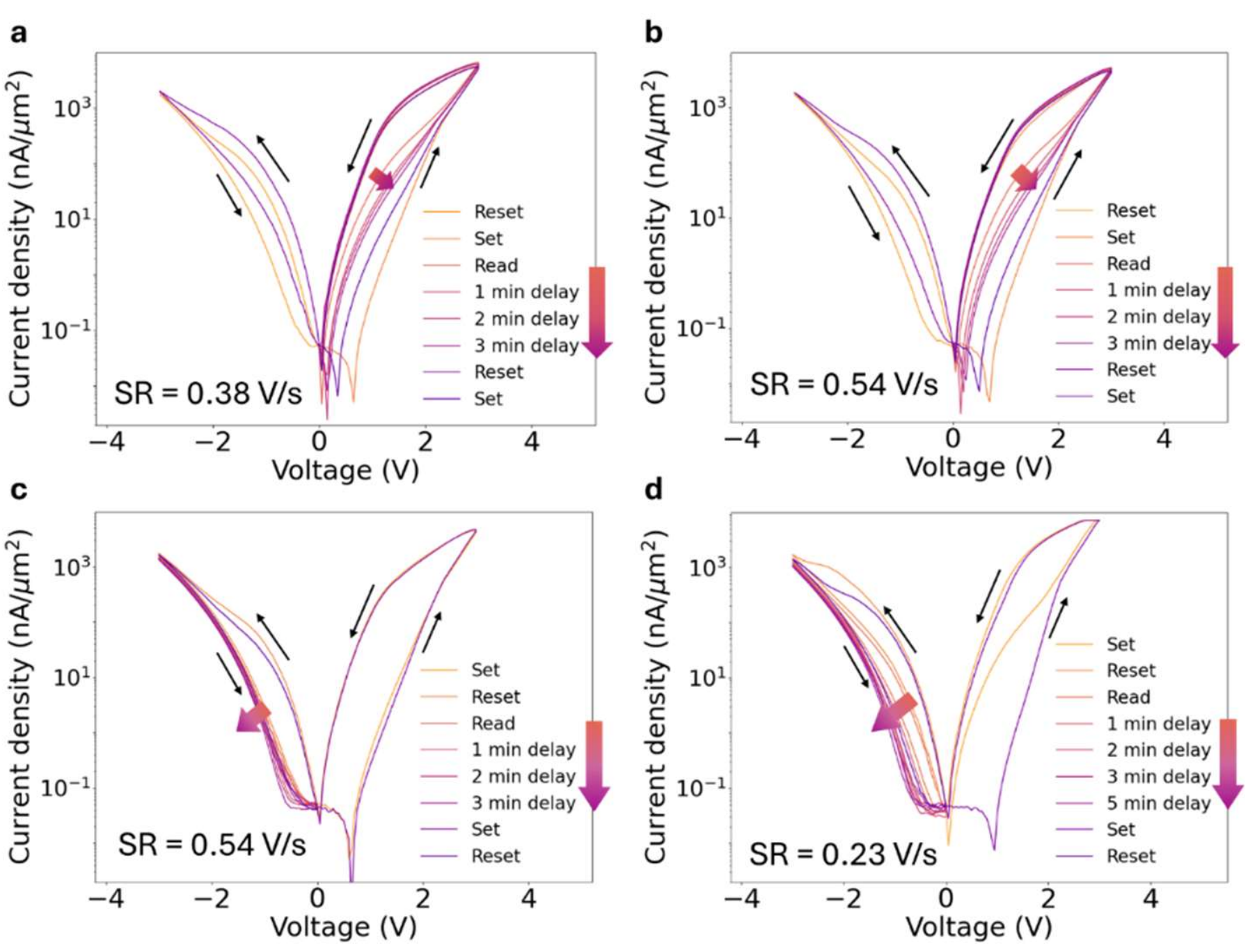

*Figure S5: Rate-dependent resistive switching at different sweep rates, as a function of delay time. (a) Resistive switching at 0.6 V/s, with delay times ranging between 0 and 3 minutes between measurements in positive polarity. (b) Resistive switching at 0.41 V/s, with delay times ranging between 0 and 3 minutes between measurements in positive polarity. (c) Resistive switching at 0.6 V/s, with delay times ranging between 0 and 3 minutes between measurements in negative polarity. (d) Resistive switching at 0.25 V/s, with delay times ranging between 0 and 5 minutes between measurements in negative polarity. In all cases, the black arrows represent the direction of the triangular sweeps (i.e. the direction of resistive switching), and the graded arrows represent the behaviour under an increasing delay time between measurements.*

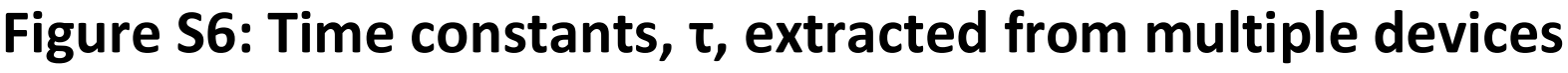

## Figure S6: Time constants, τ, extracted from multiple devices

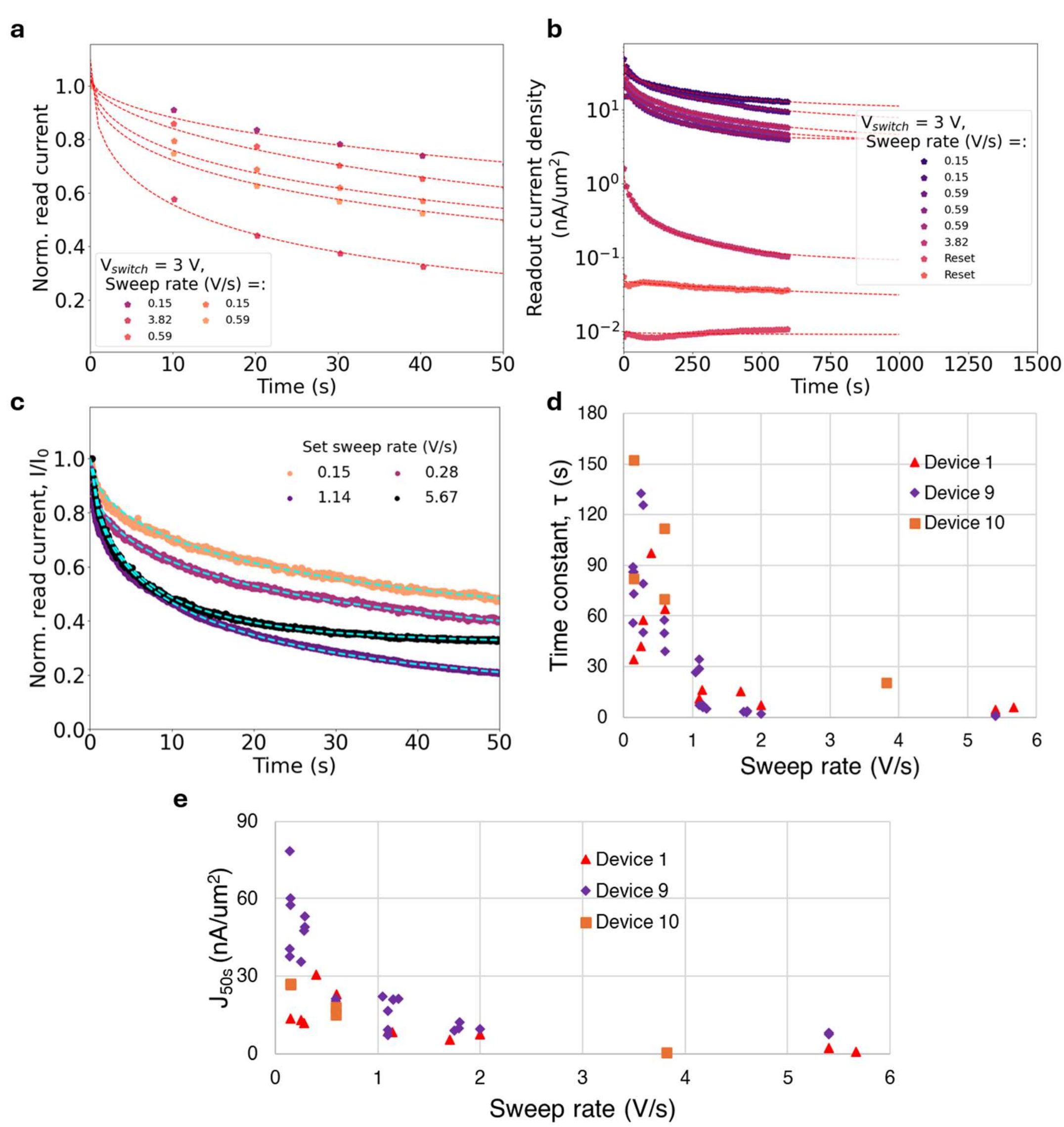

*Fig S6: (a) Retention data measured with pulsed readout on Device 10, every 10 s, for 50 s, for a series of Set sweep rates. (b) The full dataset when measuring up to 600 s (10 minutes), fitted*

*with a stretched exponential function (red dashed lines). (c) The original data from the main manuscript, reproduced here for easy visual comparison. (d) Extracted time constants vs. Set sweep rate for three different devices. (e) Measured read current density at 50 s ($J_{50s}$) vs. Set sweep rate for three different devices.*

**Figure S7: Full switching curves with varying $V_{set}$/$V_{reset}$**

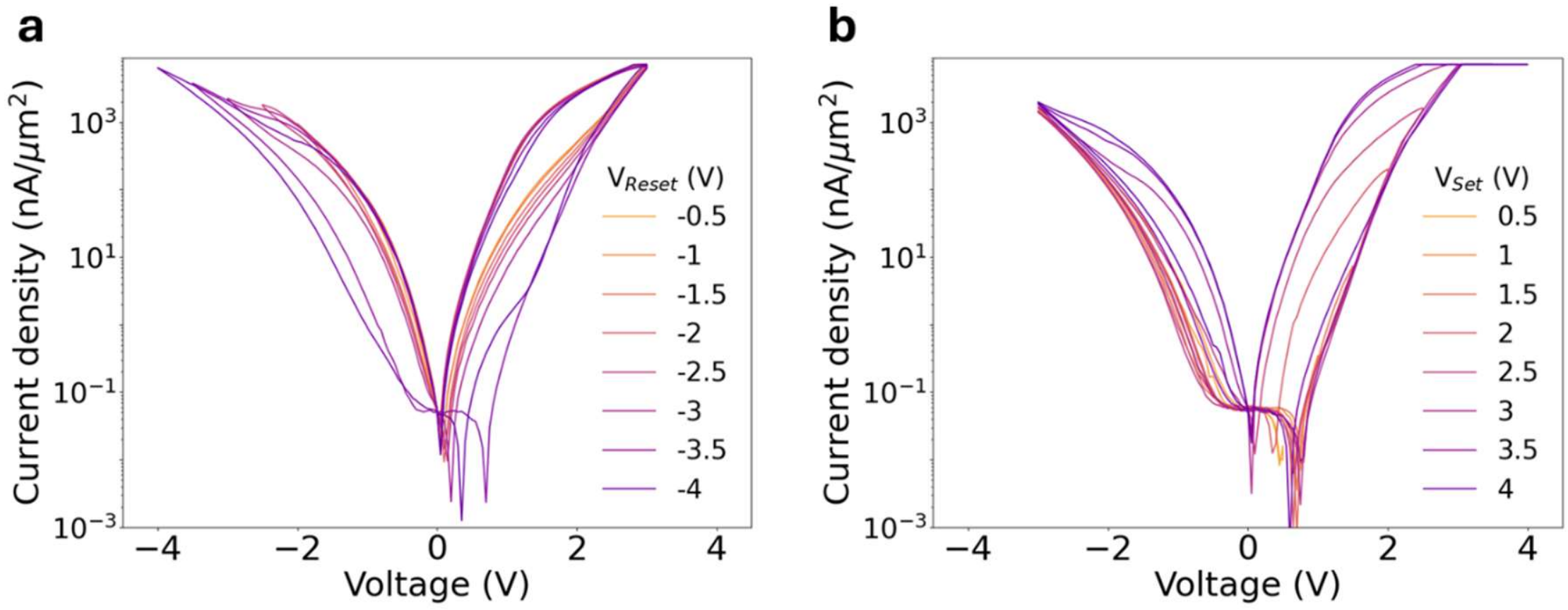

*Figure S7: (a) Switching curves measured on Device 1 when the reset voltage is decreased in steps of -0.5 V. The device is initially programmed with a 3 V positive pulse. Then, the reset voltage is incrementally decreased, with a 3 V positive pulse after each reset pulse. (b) Switching curves measured on Device 1 when the set voltage is increased in steps of 0.5 V. The device is initially programmed with a- 3 V negative pulse. Then, the set voltage is incrementally increased, with a -3 V negative pulse after each set pulse*

**Figure S8: Repeatability of redox signal**

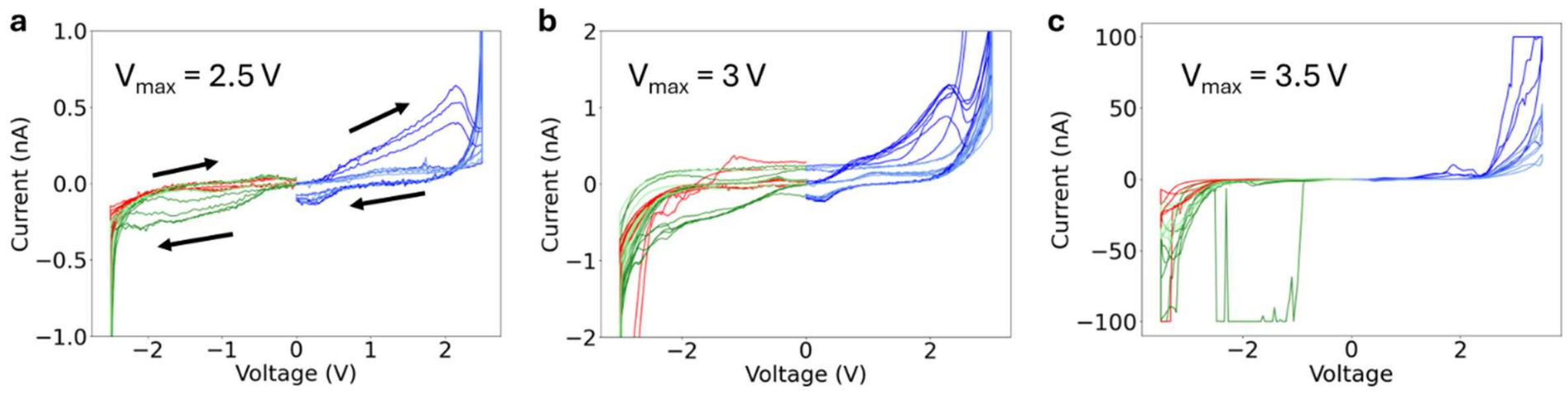

*Figure S8: Repeated switching in the soft breakdown regime. (a, b) An ionic current can be observed which grows with repeated cycling then saturates for a given voltage. It is bipolar, meaning that an opposite polarity voltage must be applied to reverse the current. A unipolar redox signal is observed on each pulse which is constant with cycles (red/light blue curves). (c) As the maximum voltage is increased, the device reaches a filamentary regime. Black arrows in (a) indicate the switching direction: note that it is opposite to the resistive switching behaviour, and although the peak behaviour resembles ferroelectricity, that can be ruled out since the integrated charge is orders of magnitude too high for polarisation switching.*

**Table S2: Parameters for calculation of diffusion coefficient**

| Parameter | Value |
|---|---|
| n | 1 *(1 electron transferred)* |
| A | 4.1e$^{-7}$ cm$^2$ *(see Table S1)* |
| C | 0.004 mol/cm$^3$ *(see below)* |

The molar concentration of $Cu^+$, *C*, can be calculated from the CCPS unit cell volume (0.785 nm$^3$), the expected number of Cu per unit cell (2) and Avogadro's constant (6.02e$^{23}$ mol$^{-1}$): $C = \frac{1}{V_{unit} N_A}$. Then the diffusion coefficient D can be calculated from the slope of the Randles-Sevcik plot following equation (1) (main text).

**Table S3: calculated diffusion coefficients in CIPS and CCPS**

| Material | Log(σT) (Sm$^{-1}$K) | σ (Sm$^{-1}$) | D (cm$^2$/s) | C (ions/m$^3$) |
|---|---|---|---|---|
| CIPS[1] | -3.74 | 6.1x10$^{-7}$ | 2.1x10$^{-13}$ | 4.8 x10$^{27}$ |
| CCPS[1] | -4.32 | 1.6x10$^{-5}$ | 5.0x10$^{-12}$ | 5.2 x10$^{27}$ |
| CIPS[2] | - | 2.7x10$^{-7}$ | 9.1x10$^{-14}$ | |
| CIPS[3] | - | 1.0x10$^{-7}$ | 3.4x10$^{-14}$ | |

Ionic conductivity σ is extracted from the references indicated and the diffusion coefficient D is then calculated from the Nernst-Einstein equation:

$$\sigma = \frac{q^2 c}{k_B T} D$$

Where q is the electrical charge of the ions (here the charge of one electron), c is the number of ions per unit volume (i.e. $Cu^+$ in CuCrP2S6, here we assume 5 x10$^{27}$ ions/m$^3$), $k_B$ is Boltzmann's constant and T is the temperature (298K).

(1) Maisonneuve, V.; Reau, J. M.; Dong, M.; Cajipe, V. B.; Payen, C.; Ravez, J. Ionic Conductivity in Ferroic CuInP2S6 and CuCrP2S6. *Ferroelectrics* **1997**, *196* (1), 257–260. https://doi.org/10.1080/00150199708224175.

(2) Dziaugys, A.; Banys, J.; Macutkevic, J.; Vysochanskii, Y. Anisotropy Effects in Thick Layered CuInP2S6 and CuInP2Se6 Crystals. *Phase Transit.* **2013**, *86* (9), 878–885. https://doi.org/10.1080/01411594.2012.745533.
(3) Zhou, Z.; Zhang, J.-J.; Turner, G. F.; Moggach, S. A.; Lekina, Y.; Morris, S.; Wang, S.; Hu, Y.; Li, Q.; Xue, J.; Feng, Z.; Yan, Q.; Weng, Y.; Xu, B.; Fang, Y.; Shen, Z. X.; Fang, L.; Dong, S.; You, L. Sliding-Mediated Ferroelectric Phase Transition in CuInP2S6 under Pressure. *Appl. Phys. Rev.* **2024**, *11* (1), 011414. https://doi.org/10.1063/5.0177451.

## Figure S9: Bipolar filament formation

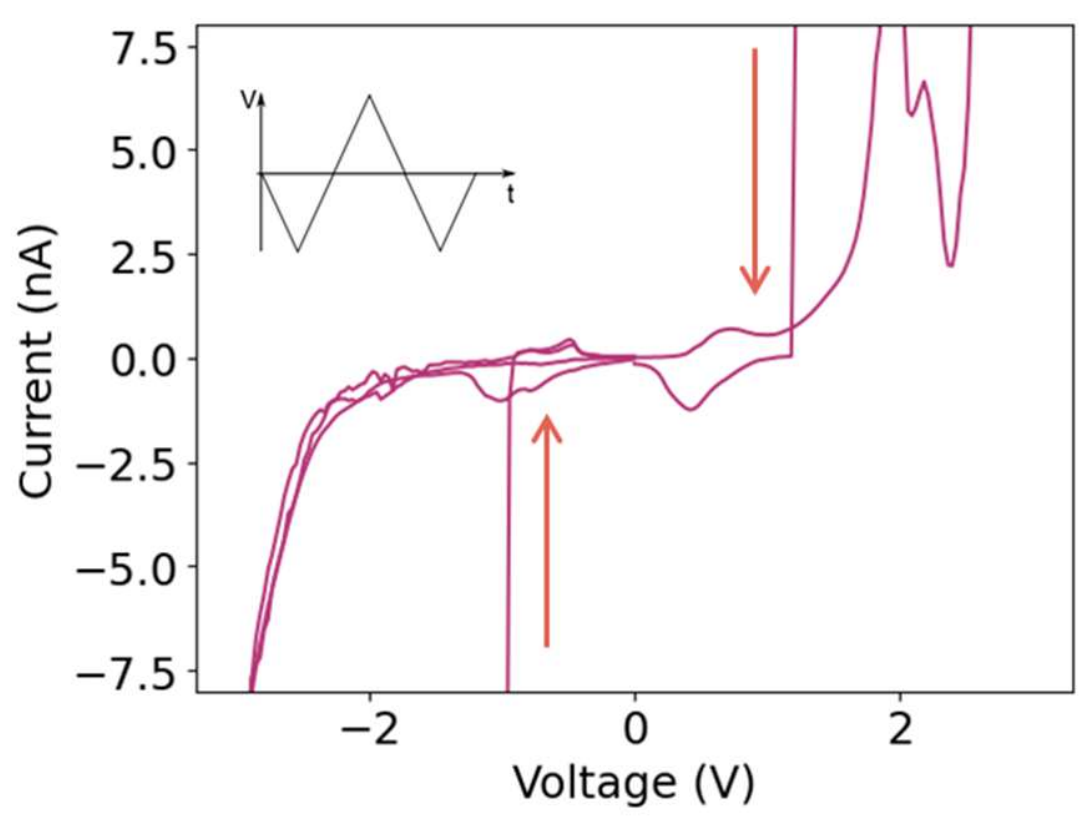

*Figure S9: Bipolar filament formation in Device 4. The device was measured with the pulse train shown inset. The first negative pulse does not show filamentary behaviour, but subsequently filaments are formed with both positive and negative pulses. The red arrows indicate the filament dissolution, which in both case happens before the redox couple peaks associated with* $Cu^0 - e^- \rightarrow Cu^+$.

## Figure S10: switching filaments On/Off with high/low voltages

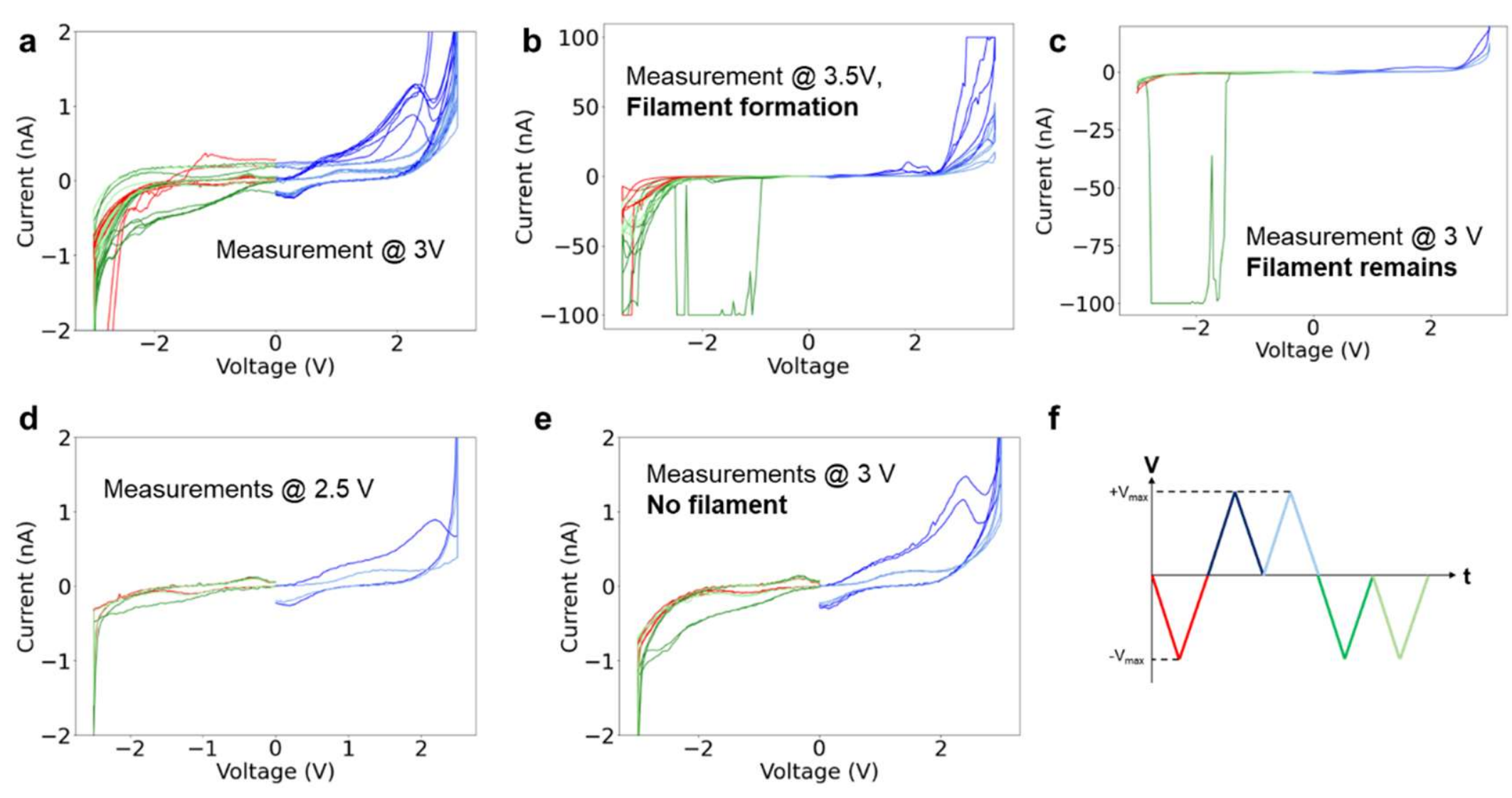

*Figure S30: filamentary switching in CCPS. (a) and (b) show the increase of the cycling voltage from 3 V to 3.5 V and the onset of filament formation. In (c), when the voltage is lowered back to 3 V, the filaments remain. Upon decreasing the voltage further, to 2.5 V, as shown in (d), no more filaments can be observed. Ramping the voltage back up to 3 V in (e), again no filamentary behaviour is observed. The pulse train applied to the device in each measurement is shown in (f). Although the process shows a higher stochasticity than represented here, this indicates that control over the formation of filaments can be achieved through the voltage applied. Note that a time-dependent dissolution of the filaments can not be ruled out here and was not investigated in this study.*

## Figure S11: Redox peaks on a device with Pt top electrode, measured in vacuum

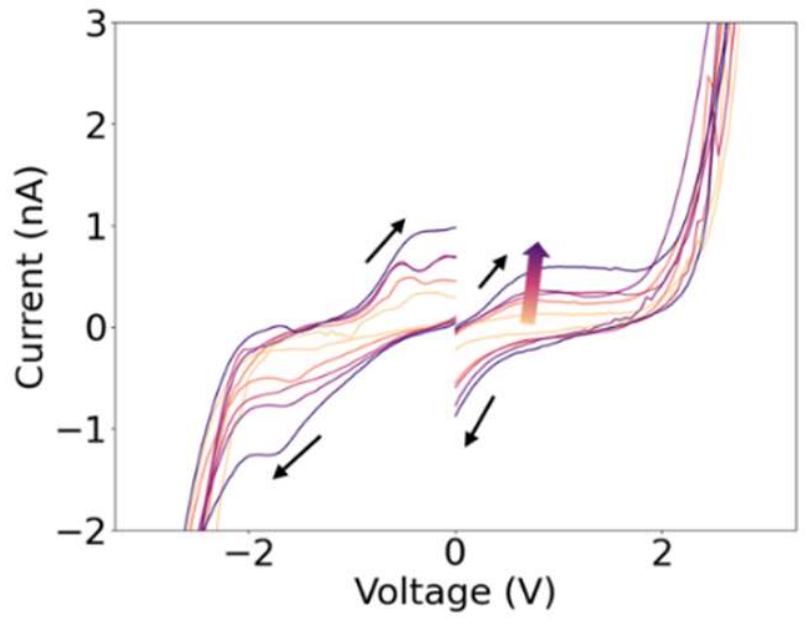

*Figure S4: Redox signatures measured on Pt/CCPS/Au devices under vacuum (0.12 mbar). The black arrows indicate the measurement direction, while the graded arrow represents an increasing sweep rate. There is a strong current discontinuity at 0 V. This is because the electrode at which voltage is being applied (and current measured) is kept constant in the measurement, but the return sweep on the positive (negative) pulse corresponds to a decreasing (increasing) voltage on the electrode, respectively, and therefore a reversal of the sign of the redox current. In the system with Pt and Au electrodes, the redox potentials are shifted with respect to the case of Cr and Au electrodes shown in the main text, leading to a redox peak appearing around 0 V. Note also that the signal is almost continuous if the absolute of the current is plotted.*

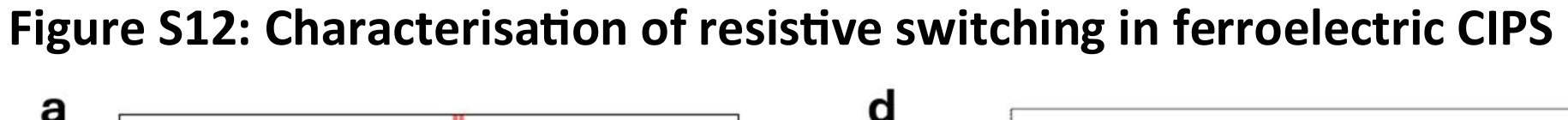

**Figure S12: Characterisation of resistive switching in ferroelectric CIPS**

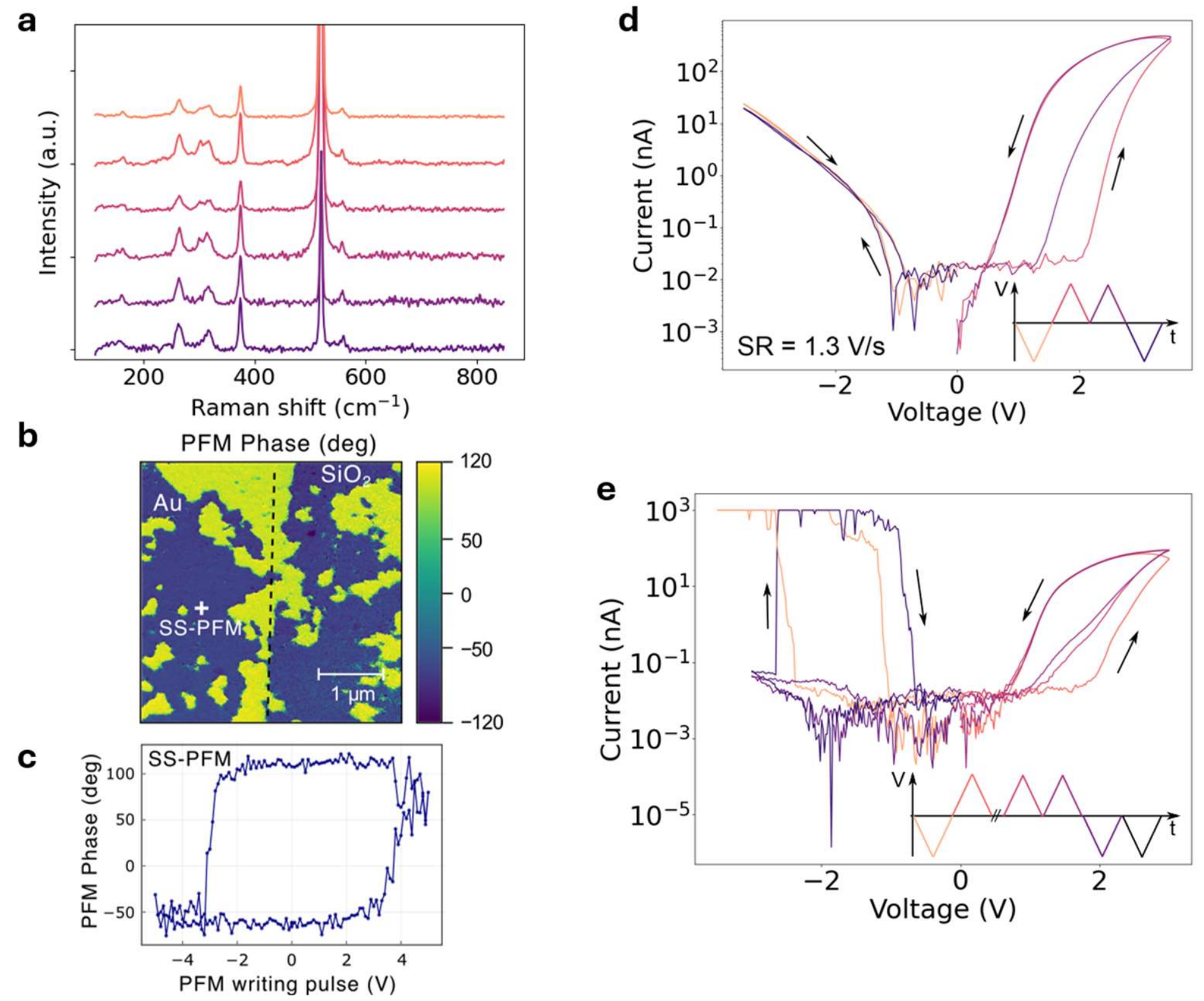

*Figure S5: Characterisation of resistive switching in ferroelectric CIPS. (a) Typical Raman spectra measured on CIPS (on-device) on different parts of the flake; the thickness is nominally equal for each measurement. (b,c) EFM phase map and SS-PFM phase hysteresis loop on CIPS on Au, showing clear domain formation and ferroelectric hysteresis. (d) Resistive switching behaviour measured on Device 7. Depending on the sweep rate, the width of the hysteresis changes. (e) Concurrent resistive switching and filament formation in CIPS. The filament dissolution voltage shifts in line with the resistive switching behaviour, since the overpotential for re-oxidation of $Cu^+$ is affected by ion migration which modifies the internal field.*

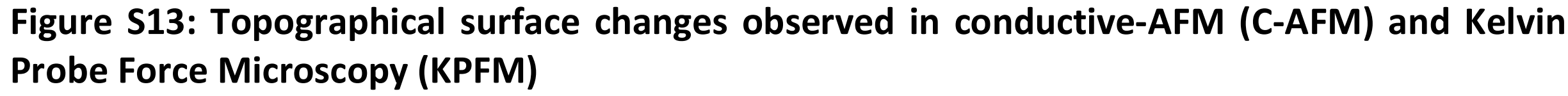

**Figure S13: Topographical surface changes observed in conductive-AFM (C-AFM) and Kelvin Probe Force Microscopy (KPFM)**

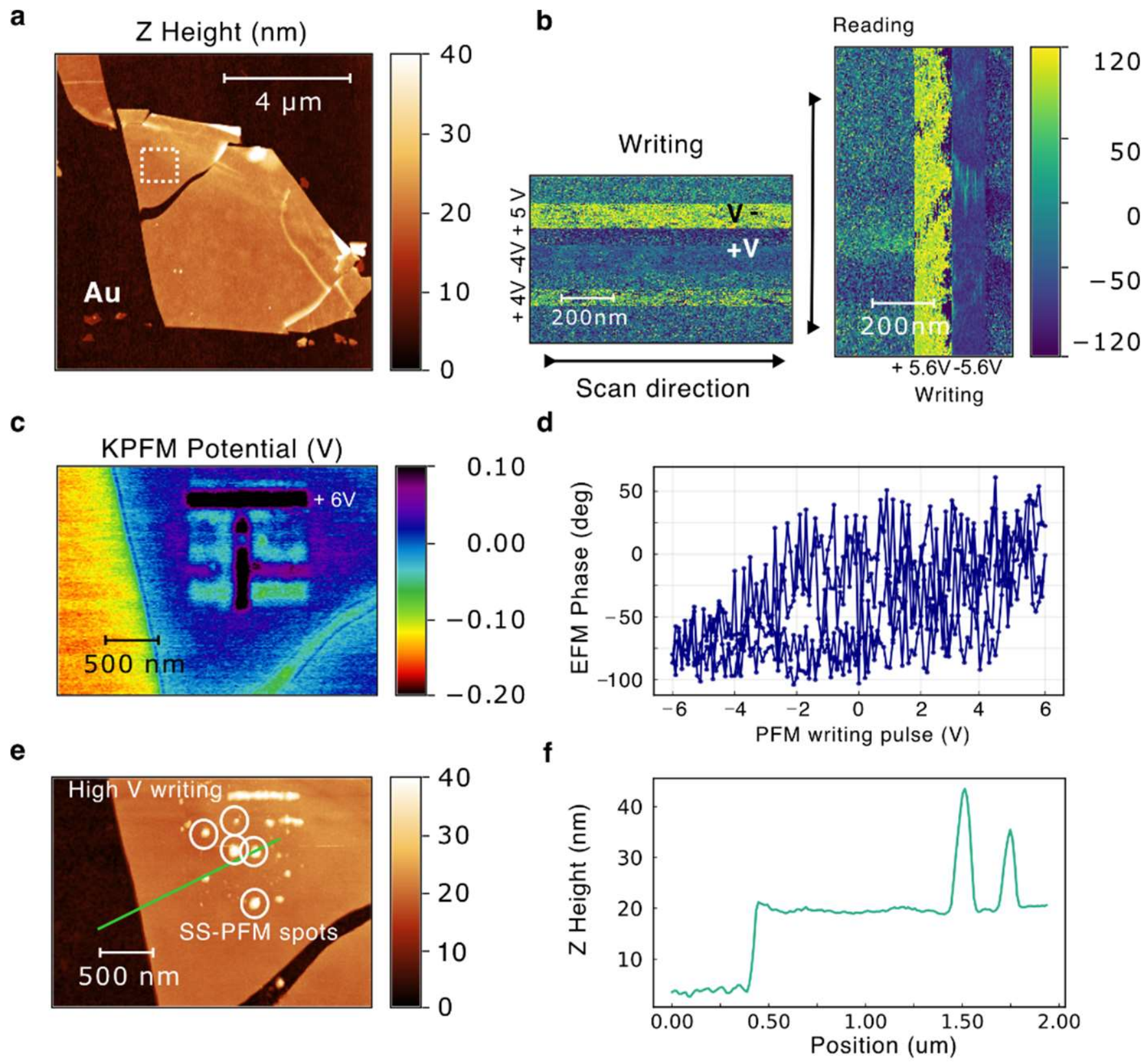

*Figure S6: Additional KPFM measurements and formation of topographical features on CCPS. (a) Z-height map of CCPS flake (≈25 nm) on Au; (b) On the left EFM phase maps of voltages written to the flake scanning horizontally the area indicated in (a), on the right same area scanned vertically, first for reading if any remanent polarization change was present and then writing with higher voltages. Equivalent to box-in-box PFM procedure, but here the amount of scanned area is reduced in order to reduce measurement time and resulting damage. (c) KPFM potential map of the structures remained after the scans in (b) and other SS-PFM acquisition. Negative writing voltages lead to a positive KPFM potential, and vice versa. The written structures could be read-out in Ar atmosphere but were unstable in air, and indeed the change in surface potential could not be read 24 hours after the sample stayed in ambient atmosphere. (d) SS-PFM phase showing absence of 180º phase shift (i.e. no ferroelectric switching occurred); (e) Z-height map after measurements. Topographical surface protrusions could be observed on the surface after SS-PFM measurements or after scanning at high voltages. The green line corresponds to a Z-height linescan extracted and plotted in (f), which show that these protrusions have a height of up to 20 nm. We attribute these protrusions to the formation of Cu clusters at the sample surface when exposed to high voltages or long scanning times.*

## Figure S14: Resistive switching on additional devices

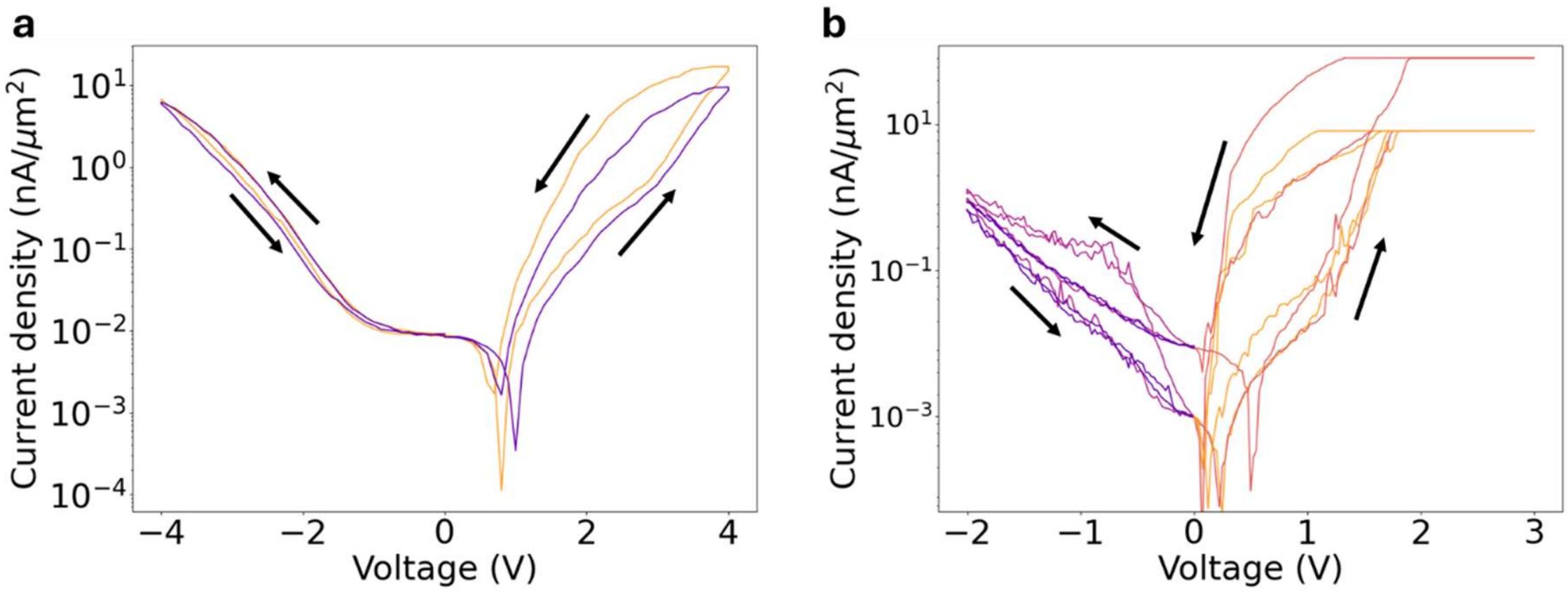

*Figure S7: Resistive switching on other CCPS-based devices. (a) Device 8 (160 nm CCPS) and (b) Device 6 (13 nm CCPS). These results indicate that the resistive switching behaviour occurs in a wide range of thicknesses of CCPS.*